\documentclass{aa}

\usepackage{newtxtext}%,newtxmath}
\usepackage[T1]{fontenc}

\DeclareRobustCommand{\VAN}[3]{#2}
\let\VANthebibliography\thebibliography
\def\thebibliography{\DeclareRobustCommand{\VAN}[3]{##3}\VANthebibliography}

\usepackage{graphicx}
\usepackage{natbib}
\usepackage{amssymb}
\usepackage{amsmath}
\usepackage{xspace}
\usepackage{dirtytalk}
\usepackage{placeins}
\usepackage[dvipsnames]{xcolor}
\usepackage{comment}
\usepackage{multirow}

\usepackage[normalem]{ulem}
\usepackage{soul}
\usepackage[varg]{txfonts}

\newcommand{\gaga}{$\gamma\gamma$}

\newcommand{\jm}[1]{{ #1}} %\textcolor{BrickRed}{{
\newcommand{\rb}[1]{{ #1}} %red \color{blue}
\newcommand{\is}[1]{{ #1}} %green \color{Plum}

\defcitealias{2022MNRAS.516..492B}{Paper I}
\defcitealias{Paper2}{Paper II}

\graphicspath{{./}{figures/}}

\begin{document}
%
% JOURNALS
%
\providecommand{\nat}{Nature}  
\providecommand{\aanda}{A\&A}  %{\textit{Astronomy \& Astrophysics}}
\providecommand{\aaps}{A\&AS}  %{\textit{Astronomy \&
                           % Astrophysics Supplement series}}
\providecommand{\aap}{A\&A}  %{\textit{Astronomy \& Astrophysics}}
\providecommand{\aapr}{A\&A Rev.}  %{\textit{Astronomy \& Astrophysics}}
\providecommand{\aj}{AJ}      %{\textit{Astronomical Journal}}
\providecommand{\apj}{ApJ}      %{\textit{Astrophysical Journal}}
\providecommand{\apjl}{ApJL}   %{\textit{Astrophysical Journal Letters}}
\providecommand{\apjs}{ApJS}   %{\textit{Astrophysical Journal Supplement}}
\providecommand{\mnras}{MNRAS} %{\textit{Monthly Notices of the R.A.S.}}
\providecommand{\memras}{Mem.~RAS} %{\textit{Mem.~of the R.A.S.}}
\providecommand{\newa}{NewA}   %{\textit{New Astronomy}}
\providecommand{\na}{NewA}
\providecommand{\jcp}{JCP}     %{\textit{Journal of Computational Physics}}
\providecommand{\rmxaa}{RMxAA} %{\textit{Revista Mexicana de Astronom\'ia y Astrof\'isica}}
\providecommand{\pasj}{PASJ}   %{\textit{Publications of the Astronomical Society of Japan}}
\providecommand{\pasp}{PASP}   %{\textit{Publications of the Astronomical Society of Pacific}}
\providecommand{\apss}{AP\&SS} %{\textit{Astrophsycis \& Space Sciences}}
\providecommand{\araa}{ARA\&A} %{\textit{Annual Reviews of Astronomy \& Astrophysics}
\providecommand{\bain}{Bull.~Astron.~Inst.~Netherlands} % {\textit{Bulletin Astronomical Institute of the Netherlands}}
\providecommand{\physrep}{Physics Reports}
\providecommand{\ssr}{Space Science Reviews}
\providecommand{\pre}{Physical Review E}
\providecommand{\jgr}{Journal of Geophysics Research}
\providecommand{\prl}{Phys.~Rev.~Lett.}% Physical Review Letters
\providecommand{\prd}{Phys.~Rev.~D}%
     % Physical Review D

\title{Interacting supernovae and where to find them}

\author{R. Brose$^{1,2,3}$\thanks{E-Mail: robert.brose@desy.de} 
    I. Sushch$^{4,5,6,7,8}$
    J. Mackey$^{3}$
    M. Arias$^{9}$}
\institute{
Institute of Physics and Astronomy, University of Potsdam, 14476 Potsdam-Golm, Germany
\and
School of Physical Sciences and Centre for Astrophysics \& Relativity, Dublin City University, Glasnevin, D09 W6Y4, Ireland.
\and
Astronomy \& Astrophysics Section, School of Cosmic Physics, Dublin Institute for Advanced Studies, DIAS Dunsink Observatory, Dublin D15 XR2R, Ireland
\and
Centro de Investigaciones Energ\'eticas, Medioambientales y Tecnol\'ogicas (CIEMAT), E-28040 Madrid, Spain
\and
Gran Sasso Science Institute, Via F.Crispi 7, 67100 L’Aquila, Italy
\and
INFN-Laboratori Nazionali del Gran Sasso, Via G. Acitelli 22, Assergi (AQ), Italy
\and
Centre for Space Research, North-West University, 2520 Potchefstroom, South Africa
\and
Astronomical Observatory of Ivan Franko National University of Lviv, Kyryla i Methodia 8, 79005 Lviv, Ukraine
\and
ASTRON - Netherlands Institute for Radio Astronomy, Oude Hoogeveensedijk 4, 7991 PD Dwingeloo, The Netherlands}

\date{Received ; accepted}
\authorrunning{R. Brose et al.}

\abstract
%Ever since the discovery of the Galactic cosmic rays, the question is which objects are able to accelerate particles to the relevant energies of at least a few PeV. \rb{Lately,} only the very early stages of remnants evolving in extremely dense circumstellar media are considered \rb{as favorable targets but have yet to be detected at gamma ray energies}.}
%{Current models for particle acceleration in very young remnants assume an oversimplified structure of the circumstellar material by implying smooth, freely expanding winds. However, the stars can undergo unsteady episodes of extreme mass-loss prior to explosion or have winds interacting with the intense photon fields of their sibling stars. In both cases, dense circumstellar shells can be formed in the star's vicinity that the supernova shocks will reach within a few days to years.}
{Early interaction of supernova blast waves with circumstellar material has the potential to accelerate particles to PeV energies, although this has not yet been detected.  Current models for this interaction assume the blast wave expands into a smooth, freely-expanding stellar wind, although multiwavelength observations of many supernovae do not support this assumption.}
{We extend previous work by considering blast waves expanding into complex density profiles consisting of smooth winds with dense circumstellar shells at various distances from the progenitor star.  We aim to predict the gamma-ray and multiwavelength signatures of circumstellar interaction.}
{We used the \textsc{PION} code to model the circumstellar medium around Luminous Blue Variables including a brief episode of enhanced mass-loss and to simulate the formation of photoionization-confined shells around Red Supergiants. Consequently, we used the time-dependent acceleration-code RATPaC to study the acceleration of cosmic rays in supernovae expanding into these media and to evaluate the emitted radiation (both thermal and non-thermal) across the whole electromagnetic spectrum.}
{We find that the interaction with the circumstellar shells can significantly boost the gamma-ray emission of a remnant, with the emission peaking weeks to years after the explosion when \gaga-absorption has reduced to negligible levels. The peak luminosity for Type-IIP and Type-IIn remnants can exceed the luminosity expected for smooth winds by orders of magnitude. For Type-IIP explosions, the light-curve peak is only reached years after the explosion, when the blast wave reaches the circumstellar shell. We evaluate the  multiwavelength signatures expected from the interaction of the blast wave with a dense circumstellar shell from radio, over optical, to thermal X-rays.}
{We identify high-cadence optical surveys and  continuous monitoring of nearby SN in radio and mm wavelengths as the best-suited strategies for identifying targets that should be followed-up by gamma-ray observatories.
We predict that gamma-rays from interaction with dense circumstellar shells may be detectable out to a few Mpc for late interaction, and tens of Mpc for early interaction.}
{}


\maketitle


\section{Introduction}

%\rb{HOW TO MAKE THIS VERY SHORT? WE HAVE PAPER II, WHICH DISCUSSES THIS ALREADY...}
%\is{I WOULD MAKE A DIFFERENT FOCUS HERE - TALK ABOUT GAMMA-RAY DETECTION OF SUPERNOVAE INSTEAD OF PARTICLE ACCELERATION IN SNRS. So, just saying a few general sentences about how SNRs are cosmic-ray accelerators which is proved through detection of gamma-ray emission, then saying that most efficiently they should accelerate at the very beginning (even beyond PeV) and cite two previous papers and then directly move to the overview of gamma-ray observations of SNe and what are the problems there.}

For almost hundred years supernova remnants (SNRs) have been extensively studied and discussed as \rb{potential} major sources of Galactic Cosmic Rays (CRs) with energies up to the knee feature in the CR spectrum at a few PeV\citep[see e.g.][]{1934PNAS...20..259B, 1987PhR...154....1B}. Advances in gamma-ray observations of historical SNRs such as Tycho and Casiopeia A, that were initially expected to be effective particle accelerators, show cutoff energies considerably below a PeV \citep{2020ApJ...894...51A}. Based on theoretical considerations, young remnants a few years after the supernova (SN) explosion moved into focus and current research investigates whether or not these objects can reach PeV energies \citep{2018MNRAS.479.4470M, 2020APh...12302492C, 2020MNRAS.494.2760C, 2021ApJ...922....7I, 2022MNRAS.516..492B, 2023ApJ...958....3D, Paper2, 2025arXiv250502523S}. 

Essential to answering this question would be a detection of the gamma-ray signatures associated with efficient particle acceleration, but so far only radio emission shows clear evidence for non-thermal emission from relativistic particles \citep{2021ApJ...908...75B} although \citet{2025arXiv250508946D} argues that the X-ray luminosity of some Type Ib/c and IIP SNe are too large to be explained by thermal emission and so a non-thermal component should be present.
Theoretically, most promising scenarios
%the SNRs most promising for reaching the highest energies 
require exceptionally dense circumstellar medium (CSM) \citep{2013MNRAS.431..415B}, and phenomenological models suggest that those events might be detectable out to distances of hundreds of Mpc \citep{MurThoLac11, MurThoOfe14}. 
Work that also accounted for the \gaga-absorption of the gamma-rays by interaction with photons from the SN photosphere obtained considerably lower horizon estimates, at Mpc scales \citep{2009A&A...499..191T}. Many works followed that, with more detailed descriptions of the acceleration and \gaga-absorption, estimated similar detection horizons for those young objects \citep[and references therin]{2020MNRAS.494.2760C, 2022MNRAS.511.3321C, 2022MNRAS.516..492B}.

Despite intensive experimental effort, a detection of gamma-rays from a very young SNR has yet to be reported. There are hints of a transient $\gamma$-ray signal from \textit{Fermi-LAT} data overlapping with the positions of SN 2004dj, a Type-IIP explosion \citep{2020ApJ...896L..33X}, and SN iPTF14hls \citep{2018ApJ...854L..18Y}, at distances of $3.5\,$Mpc and $156\,$Mpc, respectively. Additionally, two more SN candidates have been found in a transient search throughout the \textit{Fermi-LAT} archive \citep{2021MNRAS.tmp.1282P}. \textit{Fermi-LAT} upper limits towards the nearby SN 2023ixf\footnote{The estimated distance is 6.85Mpc.} have been used to constrain the acceleration efficiency to values $<1$\% \citep{2024A&A...686A.254M}.
Similarly the nearby SN 2024ggi has only  \textit{Fermi-LAT} upper limits reported \citep{2024ATel16601....1M}.
A systematic search for gamma-rays associated with Type-IIn SNe - the subclass with the highest known CSM densities - showed no signal for individual SNe nor the combined sample of SNe \citep{2015ApJ...807..169A}. Despite dedicated observing campaigns and searches, there is also no detection of very-high energy (VHE) gamma-ray emission from any core-collapse SNe \citep{2019A&A...626A..57H, Abdalla:2021Mg}.

In contrast to the lack of detection of SNe, the much less energetic novae are now established as GeV \citep{2014Sci...345..554A} and TeV \citep{HESS2022_RSOph, 2022NatAs...6..689A} sources of gamma-rays, implying efficient particle acceleration in the first few days post-explosion.
The physical processes and conditions of blast-wave propagation in core-collapse SNe are very similar to recurrent novae in many respects, if energy and ejecta mass are scaled up by a few orders of magnitude.
This reinforces the consensus that GeV and TeV emission will be detected from SNe if they explode sufficiently close to Earth.

A shortcoming of all the theoretical work cited above is that it is assumed that the progenitor star had a constant mass-loss rate on the relevant timescales prior to the explosion, resulting in a smooth wind-profile where  gas density $\rho(r)\propto r^{-2}$ as a function of distance, $r$, from the star.
However, from the stellar evolution point of view, this is a strong assumption and %does not hold in reality.
may not be valid in many cases.
In the late evolutionary stages massive stars may undergo rapid changes in radius with associated changes in mass-loss rate and wind speed \citep[e.g.][]{Lan12}.
Evidence is increasing that especially Type-IIP explosions show enhanced mass-loss just prior to explosion \citep{2022ApJ...924...15J, 2025arXiv250504698J}.
Recent studies of radio emission from core-collapse SNe \citep{2025arXiv250506609M} also support the conclusion that a region of very dense CSM exists around many SN progenitors, which further challenges the above-mentioned theoretical models.

The matter is further complicated by Luminous Blue Variable stars (LBVs) - the likely progenitors of Type-IIn SNe - undergoing giant eruptions where a few solar masses of material are ejected on a timescale of a year, expanding with $\sim100$\,km\,s$^{-1}$, the most famous example being $\eta$ Carinae \citep{Smi14}.
Red supergiants (RSGs) on the other hand might feature external photoionization of their wind from nearby hot stars that may shock and partially confine the RSG wind into dense shells that exhibit large compression ratios with respect to the freely-expanding wind \citep{2014Natur.512..282M}.
Such photoionization-confined shells should have a typical distance of 0.01-1\,pc from the RSG, depending on wind density and the intensity of the external radiation field.

As a result of the structured nature of the CSM, the SNR blast wave may only interact with the dense CSM with a considerable delay.
For example dense shells at $r\sim0.01-1$\,pc will trigger late-time circumstellar interaction years after explosion, as detected in infrared surveys \citep{2013AJ....146....2F}.
This challenges existing theoretical models and observational strategies: Imaging Air Cherenkov Telescopes (IACTs) usually try to observe SNe within the first weeks after explosion and \textit{Fermi-LAT} analysis often focuses on the emission within the first year after explosion, so that later emission can easily be missed. A notable exception here is the Type-IIP SN 2004dj, where the reported gamma-ray signal appears $\sim5$ years post-explosion \citep{2020ApJ...896L..33X}, roughly coinciding with the time at which a blast wave would reach the typical location of a photoionization-confined shell. However, if CSM interactions are common, they will induce a late time brightening in gamma rays, accompanied by signatures in other wavelengths. It is important to develop trigger strategies for current and next generation gamma-ray telescopes that take this possibility into account.

This work builds on our previous publications, \citet[][hereafter \citetalias{2022MNRAS.516..492B}]{2022MNRAS.516..492B} and \citet[][hereafter \citetalias{Paper2}]{Paper2}. In \citetalias{2022MNRAS.516..492B}, we studied particle acceleration and non-thermal emission from SNRs associated with explosions of LBV and RSG stars for explosions into smooth $1/r^2$ density profiles appropriate for stellar winds without variability. There, we developed the revenant descriptions for the emission and absorption processes relevant for these very young SNRs.
In \citetalias{Paper2}, we studied how the interaction of a SN blast wave with dense shells in Type-IIn explosions (associated with LBV progenitors) affects the maximum energy achievable in those cases. We found that the interaction with the dense shells can boost the maximum energy beyond a PeV if the interactions happen early enough, illustrating the importance of a more realistic description of the CSM around massive stars. 

In this work, we focus on the multiwavelength signatures that arise from the interaction of the blast waves with these dense shells. While \citetalias{Paper2} was solely focused on LBV progenitors, we additionally consider RSG progenitors in this work, where the dense photoionization-confined shells may not affect the achievable maximum energy but can significantly alter the multiwavelength lightcurves.

Section~\ref{sec:methods} describes the model setup and introduces our assumptions for density and magnetic field of pre-supernova CSM. We also further discuss the implications of certain aspects of our numeric simulations and the assumptions that they rely upon. In section~\ref{sec:results} we describe the emission signatures from gamma-rays to radio and the horizon out to which supernovae can be detected by current and future observatories, followed by a revised observation strategy (section~\ref{sec:strategy}) based on our findings. Section~\ref{sec:conclusions} presents our conclusions.

\section{Model setup}
\label{sec:methods}

The numerical setup of simulations as well as basic assumptions and equations that are solved are identical to \citetalias{Paper2}. We use the RATPaC code to combine a kinetic treatment of the CRs with a thermal leakage injection model, a fully time-dependent treatment of the magnetic turbulence, and a \textsc{PLUTO}-based hydrodynamic calculation \rb{\citep[][]{Telezhinsky.2013a, 2016A&A...593A..20B, 2007ApJS..170..228M}}.
All of the calculations are performed assuming spherical symmetry on a 1D computational grid.
The essential ansatz to describe the evolution of the CR-distribution is solving the kinetic diffusion-advection equation for CR-transport \citep{Skilling.1975a}
\begin{align}
    \frac{\partial N}{\partial t} =& \nabla(D_r\nabla N-\mathbf{u} N)\nonumber\\
 &-\frac{\partial}{\partial p}\left( (N\dot{p})-\frac{\nabla \cdot \mathbf{ u}}{3}Np\right)+Q
\label{CRTE}\text{ , }
\end{align}
where $D_r$ denotes the spatial diffusion coefficient, $\textbf{u}$ the advective velocity, $\dot{p}$ energy losses and $Q$ the source of thermal particles.

For detailed description of the setup we refer the reader to \citetalias{2022MNRAS.516..492B} and \citetalias{Paper2}, while here we focus only on the additional assumptions made for the RSG progenitors. An in-depth discussion of the assumed microphysics of the self-generated turbulence and their impact on the results can be found in \citetalias{Paper2}.

\subsection{Circumstellar medium}
In this work, we consider LBV and RSG progenitors as the most optimistic scenarios for enhanced emission due to interaction with CSM shells. The densest CSM is expected to be formed by LBV eruptions, where the mass-loss rate can reach values as high as $1\, \mathrm{M}_\odot\,$yr$^{-1}$ for periods of a few years \citep{Smi14}. It is slowly becoming evident that RSGs might also have enhanced mass-loss rates of up to $10^{-2}\, \mathrm{M}_\odot\,$yr$^{-1}$ a few years prior to explosion \citep{2022ApJ...924...15J}. Moreover, some RSGs may feature dense photoionization-confined shells that can trap up to 35\% of the mass lost during the RSG phase of stellar evolution \citep{2014Natur.512..282M}. While Type IIn SNe that are observationally associated with LBVs constitute only $\sim 5\,$\% of the CCSN rate \citep{2023A&A...670A..48C}, RSGs are the most common SNe progenitors resulting in Type IIP explosions that comprise $\sim50\,$\% of all CCSNe \citep{2011MNRAS.412.1522S}.
\jm{It must be noted, however, that the large majority of SN IIP do not interact with dense circumstellar shells at tens of mpc scales and so the incidence of such shells should be low.}

%\rb{In this work, we consider a standard ISM composition for the CSM material, even though a heavier composition can be expected in the vicinity of massive stars. A heavier composition would not affect the light curves or luminosity considerably but could change the spectra shape of the gamma-ray emission, especially at energies below $1\,$GeV \citep{2020APh...12302490B, 2020MNRAS.497.3581D}.}

\subsubsection{LBV shell modeling}\label{sec:LBV-wind}

The simplest model for a LBV eruption is to impose a dramatically increased mass-loss rate for a short period of time.
We modelled an evolving stellar wind with \textsc{pion} \jm{\citep{2021MNRAS.504..983M}} in spherical symmetry, by simulating the expansion of a stellar wind in three phases. The first phase has mass-loss rate $\dot{M} = 10^{-4}\,\mathrm{M}_\odot\,\mathrm{yr}^{-1}$ and wind terminal velocity $v_\infty=100$\,km\,s$^{-1}$ and lasts for 1800 years; the second (outburst) phase has $\dot{M} = 1\,\mathrm{M}_\odot\,\mathrm{yr}^{-1}$ and $v_\infty=100$\,km\,s$^{-1}$ and lasts for 2 years; the third is the same as the first but we integrate for 10\,000 years so that the shell expands to about 1 parsec radius.
The density structure of the shell is then used to construct the circumstellar shell in the \textsc{RATPaC} simulations as described below in section~\ref{sec:shell-model}, where some smoothing is applied so that the shock-finding algorithm works as intended. Examples of the CSM structure can be found in section 2.1 of \citetalias{Paper2}.

\subsubsection{Photoionized shells around RSGs}\label{sec:RSG-wind}

For the photoionization-confined shell model, we consider two different cases to produce a high-mass and low-mass CSM shell.
For the high-mass shell we consider a RSG with $\dot{M} = 10^{-4}\,\mathrm{M}_\odot\,\mathrm{yr}^{-1}$ and $v_\infty=15$\,km\,s$^{-1}$, a mass-loss rate appropriate for the most massive RSGs like VY CMa \citep[e.g.][]{SmiHinRyd09}.
This wind is exposed to an external ionizing flux of $F_\gamma = 1.7\times10^{12}$\,s$^{-1}$, following the methods presented in \citet{2014Natur.512..282M} and \citet{2015A&A...582A..24M}.
The system is evolved in spherical symmetry with \textsc{pion} for 200\,kyr, during which time the RSG loses 20\,M$_\odot$ through stellar wind.
A dense, cold and thin shell forms at a radius of about 0.03\,pc in the wind, because of the D-type ionization front that forms in externally photoionized wind.
The shock is essentially isothermal and so the compression factor is determined by the minimum temperature obtained, in this case about 100\,K.
At this radius the wind density is $\rho\approx2\times10^{-20}\,\mathrm{g\,cm}^{-3}$, whereas the thin shell is more than $100$ times denser.
For the low-mass shell the RSG wind is diluted to $\dot{M} = 2\times10^{-5}\,\mathrm{M}_\odot\,\mathrm{yr}^{-1}$ and the ionizing flux to $F_\gamma = 1.7\times10^{10}$\,s$^{-1}$, resulting in a CSM shell with about 10 times less mass than the high-mass case.
\jm{The parameters used for the two photoionised shells are identical to those used for two calculations in \citet{2015A&A...582A..24M}, modelling the H$\alpha$ nebula around the star Wd1-26 in Westerlund 1.}
The shell properties are given in Table~\ref{tab:ProgenitorModels} and density profiles plotted in Fig.~\ref{fig:Shells}.
As with the LBV case, the \textsc{pion} density field is used to construct the pre-SN CSM in the RATPaC calculation as described below. 

\subsubsection{Unified shell model} \label{sec:shell-model}
In \citetalias{Paper2} we modeled the CSM density profile only for the LBVs. Here we additionally introduce shells of enhanced mass following a Gaussian distribution for the RSGs using the same prescription,
\begin{align}
 \rho(r) = \frac{M_\text{shell}}{4\pi r^2\sqrt{2\pi d_\text{shell}^2}}\exp\left(-\frac{(r-R_\text{shell})^2}{2 d_\text{shell}^2}\right)
 \text{,}\label{eq:ShellProfile}
\end{align}
where $M_\text{shell}$ is the shell mass, $d_\text{shell}$ the shell thickness and $R_\text{shell}$ the radial position of the shell. The parameters for the free winds and the corresponding shells are given in Table~\ref{tab:ProgenitorModels}.

\begin{table*}[h]
  \centering
  \caption{Parameters for the progenitor stars winds and initial remnant sizes.
  %\jm{[I would suggest removing $R_{ej}$ from this table since it is a numerical parameter and all others are physical.  Also I prefer $2\times10{30}$ instead of $2\cdot10^{30}$ but I guess this is a style choice?  I may have added some $\times$ symbols to the text\ldots]}
  }
  \label{tab:ProgenitorModels}
  \begin{tabular}{c|c c | c c c | c c c}
    %\hline
     & \multicolumn{2}{c|}{Smooth Wind} & \multicolumn{3}{c|}{SN ejecta} & \multicolumn{3}{c}{Circumstellar Shell} \\
    Model & $\dot{M} [M_\odot$\,yr$^{-1}$] & $V_\mathrm{w}$ [km\,s$^{-1}$]  & $R_\text{ej}$ [cm] & $M_\text{ej}$ [$M_\odot$] & $n$ & $M_\text{shell}$ [$M_\odot$] & $d_\text{shell}$ [mpc] & $R_\text{shell}$ [mpc]\\
    \hline
    %\hline
    %& & & & \\
    LBV 0.1 & \multirow{6}{*}{$10^{-2}$} & \multirow{6}{*}{100} & \multirow{6}{*}{$1.2\times10^{14}$} & \multirow{6}{*}{10} & \multirow{6}{*}{10} & \multirow{6}{*}{2} & \multirow{6}{*}{0.2} & 1.0\\
    LBV 0.3 &  &  &  &  &  &  &  & 3.0\\
    LBV 0.5 &  &  &  &  &  &  &  & 4.3\\
    LBV 0.75 &  &  &  &  &  &  &  & 6.2\\
    LBV 1.0 &  &  &  &  &  &  &  & 8.0\\
    LBV 2.0 &  &  &  &  &  &  &  & 15.0\\
    LBV 10 &  &  &  &  &  &  &  & 61.0\\
    RSG lowmass & $2\times10^{-5}$ & 15 & \multirow{2}{*}{$6\times10^{13}$} & \multirow{2}{*}{3} & \multirow{2}{*}{9} & 0.58 & 0.43 & 48.8\\
    RSG highmass & $9.9\times10^{-5}$ & 15 &  &  &  & 4.41 & 0.313 & 29.9\\
    %Type-Ib/c & $2.5\times10^{-6}$ & 1000 & $6\times10^{11}$ & &\\
    %\hline
  \end{tabular}
  \tablebib{\rb{The values for the winds are chosen in accordance with \cite{2014MNRAS.440.1917D} for Type-IIP and \cite{2021A&A...647A..99G} for Type-IIn SNe.}} Columns 2-3 give the mass-loss rate (2) and wind velocity (3) (assumed constant for the free wind), and colums 4-6 give the initial radius of the SN ejecta at the start of the simulation (4), the total ejecta mass of the SN (5) and the parameter $n$ determining the radial dependence of the ejecta (6). Colums 7-9 give the shell mass (7), shell thickness (8) and position of the shell (9) (see text for details).
\end{table*}

\begin{figure}
    \centering
    \includegraphics[width=0.49\textwidth]{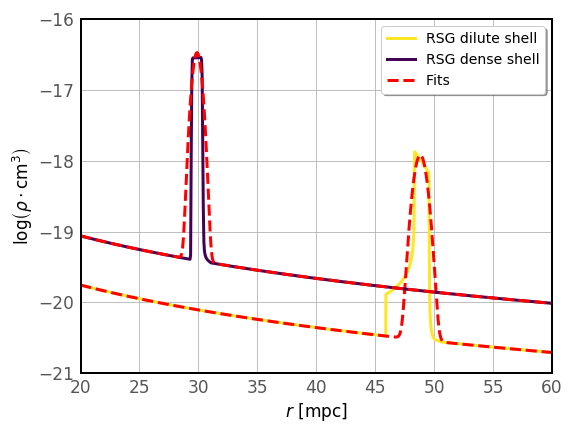}
    \caption{Density profile of the CSM for the two RSG scenarios listed in table \ref{tab:ProgenitorModels}. The low \rb{mass loss} case creates a more dilute (yellow) shell compared to the high mass-loss case (purple). The red dashed lines are the desnity-profiles adopted in our simulations.}
    \label{fig:Shells}
\end{figure}

We obtained the shell parameters by fitting the results of the stellar-wind simulations described in sections~\ref{sec:LBV-wind} and ~\ref{sec:RSG-wind} with the Gaussian profiles described  by equation~(\ref{eq:ShellProfile}). An example for the CSM structure as obtained by the \textsc{Pion} simulations for the two RSG cases is shown in Figure \ref{fig:Shells}. Figure 1 of \citepalias{Paper2} shows examples of the density profiles that were adopted for the LBV-cases.

The total mass of material in the CSM is important not only for the shock dynamics but also for absorption in the ambient medium, for instance of X-rays and radio waves. 
It has to be noted that we ignore the effects of radiative cooling in our calculations. Strictly speaking, especially around the reverse-shock, the shock will be radiative initially but the intense photon-fields would require the use of a different equation of state for the plasma. The interactions with the dense shells will push the shock-dynamics into the radiative regime for a brief period even for the forward-shock for the LBV cases with the earliest interactions.
For such cases, the equations of radiation-hydrodynamics should be solved with multi-group radiative transfer, \citep[e.g.][]{2013MNRAS.428.1020M}, which is beyond the capabilities of any code that also calculates particle acceleration and transport.
This unavoidable limitation introduces some uncertainty to our models, in particular in the spatial separation between the forward and reverse shocks, and the density of the shocked gas.
Investigating this uncertainty is planned for future work where we consider a more general equation of state for the thermal plasma.
%\rb{Also here, a part of the created intense photon fields would be reabsorbed by the plasma, so that a different equation of state would needed to be used. Resolving} the changed dynamics is clearly beyond the scope of this paper \rb{and its biggest impact can be considered our inability to predict the optical rebrightening as calculated in section \ref{sec:Optical_method}.} \is{UFFF, A REALLY GREAT PLACE FOR THE REFEREE TO GET HOOKED AND ASK WHAT IMPACT THE RADIATIVE SHOCK WOULD HAVE ON THE PARTICLE ACCELERATION... VERY RISKY}\rb{SO WHAT? '^^ DON'T SAY THAT? NOT SURE WHAT TO DO HERE THEN...}
%\is{[IS: I am not sure either. If we say this maybe we should also say that it's not clear how the radiative regime will impact acceleration and add a reference of some kind? Just to show that we are aware of this problem. Because so far it sounds like this will only impact the shock dynamics.]}\rb{I slightly rephrased that...}
%\jm{I made a suggestion above\ldots not sure what you think?  I would avoid any specific statement about what we can and cannot predict because we don't really know.  I would prefer something vague that will be investigated in future work.}
%\jm{[Is it really clear that the RS is always radiative?  Can we soften this statement such that we say it may be radiative at certain times for the simulations with the densest CSM, or do you expect it is always radiative?]}

%\is{[IS: SHOULD WE ADD THE PLOT FOR THE PROFILE WITH A SHELL AS AND ILLUSTRATION?]}}\rb{GOOD IDEA. WILL DO THAT! ALSO ADD DISCUSSION ON COOLING.}

\subsection{Supernova injection}
We inject the supernova following the methods described in \citetalias{Paper2}, with ejecta mass, $M_\mathrm{ej}$, index of the density profile of the outer ejecta, $n$, and ejecta outer radius, $R_\mathrm{ej}$, given in Table~\ref{tab:ProgenitorModels}.
In all cases we assume an explosion energy of $10^{51}$\,erg. \rb{Observed explosion energies usually lie within a factor of a few within this canonical value, but the difference arising from a different explosion energy are easily absorbed in the uncertainties of our other parameters, e.g. the shell positions.

In reality, deviations from spherical symmetry might introduce further effects, e.g. a time-offset of the interaction between the shocks and the shells in different parts of the remnant. However, considering these effects requieres a full spatial 2D treatment and is beyond the capabilities of current-generation acceleration codes like RATPaC. 
}

%[Why such a low ejecta mass for RSG progenitor - I would have thought it should be more like 10Msun?]

\subsection{Self-generation of magnetic turbulence}\label{sec:Turb}
In this work, we assess the magnetic turbulence by solving the transport equation for the turbulence spectrum in parallel to the transport equation of CRs \citep{2016A&A...593A..20B}:
%\is{IS: Ehm... this sounds like an overstatement for linear factor in the growth rate that we use... I mean, physically speaking we do not "add parts related to non-resonant mode". Like nothing in our treatment of turbulence has anything to do with non-resonant. We just make resonant instability artificially strong enough to get to the values of B-field expected for non-resonant (or even rather measured for young SNRs). I think we shuld be honest and careful here} In any case, the diffusion coefficient varies strongly in space and time and is coupled to the spectral energy-density per unit logarithmic bandwidth, $E_w$. The evolution of $E_w$ is described by
\begin{align}
 \frac{\partial E_w}{\partial t} +   \nabla \cdot (\mathbf{u} E_w) + k\frac{\partial}{\partial k}\left( {k^2} D_k \frac{\partial}{\partial k} \frac{E_w}{k^3}\right) = \nonumber\\
=2(\Gamma_g-\Gamma_d)E_w \text{ . }
+ Q
\label{eq:Turb_1}
\end{align}
Here, $\mathbf{u}$ denotes the advection velocity, $k$ the wavenumber, $D_k$ the diffusion coefficient in wavenumber space, and $\Gamma_g$ and $\Gamma_d$ the growth and damping terms, respectively \citep{2016A&A...593A..20B}. Thus, additionally to equation (\ref{CRTE}), where we describe the spatial distribution of CRs depending on their energy, we obtain the distribution and energy density of magnetic turbulence in a spatially and temporal resolved manner as well.

For details of our treatment, we refer the reader to \citetalias{Paper2}.
The turbulence description is essential in understanding the impact of the \citetalias{Paper2}, where we have an in-depth discussion on our assumptions and put them in perspective with other models in the literature.
While for \citetalias{Paper2}, the turbulence description is essential in understanding the impact of the shell interaction on the maximum energy, this is less important when considering the emission signatures which are the focus of this paper. The most profound impact of our turbulence description, and hence the acceleration time of particles, manifests itself in the delay times between the peaks of the emission in different energy bands.

\subsection{Radiation and absorption processes}\label{sec:Absorption}
To calculate non-thermal emission from the SNR we consider synchrotron radiation in the radio and X-ray energy bands and proton interactions with subsequent pion decay in the gamma-ray energy band. For our assumed electron-to-proton ratio, the VHE gamma-ray emission is about equal from Pion decay and inverse-Comption (IC) scattering for ambient densities of the order of $1\text{cm}^{-3}$ and the CMB as a target photon field \citep{2021A&A...654A.139B}. As a consequence, IC emission is negligible for the scenarios presented here, as the photosphere might only provide additional target-photons at times where the $\gamma$-ray emission is suppressed strongest by \gaga-absorption. Further, gamma-ray production on photons from the photosphere is strongly Klein-Nishina suppressed in the VHE band \footnote{\rb{The external photon field around RSGs (see also section \ref{sec:RSG-wind}) might provide additional seed-photons but is orders of magnitude below the energy density of photons from the SN-photosphere and might only get relevant at significantly later times than the ones considered here.}}.
%Gamma-ray emission produced through the inverse Compton scattering in considered scenarios appears to be negligible comparing to that from hadronic interactions and therefore was ignored.
Additionally, we model the thermal continuum X-ray emission to enable comparisons to observational data. 

The high matter- and photon-densities make it necessary to account for local absorption of X-rays by ions of heavy elements in a gas with approximately solar abundances; free-free absorption at radio-wavelengths and attenuation of gamma-rays by anisotropic photon-photon collisions.
We therefore attenuate the flux emitted from every part of the remnant by the optical depth produced by the matter that has to be transversed by the rays along the line-of sight according to 
\begin{align}
    F(x,E) &= F_0(x,E)\exp\left(-\tau(x,E)\right) \text{ , }
\end{align}
where $\tau$ is the optical depth from observer to the point $x$, and $F$ and $F_0$ are the attenuated and unattenuated photon-fluxes at a given location and energy respectively. The total flux is then obtained by summing the contributions from all parts of the remnant.
As noted above, all of our calculations assume a spherically symmetric matter distribution, and we consider an observer at large distance from the SN so that lines of sight are parallel to each other.

Further, we account for gamma-ray absorption by extragalactic background light for SNRs at large distances. The calculation of the emission and absorption processes follows the steps described in \citetalias{2022MNRAS.516..492B} and \citetalias{Paper2}. A more detailed description of the emission and absorption processes can be found in appendix \ref{sec:Ap_rad}.

% \rb{The outline of the emission and absorption processes closely follows our treatment from \citetalias{Paper2}.}

%\is{I am fine with just citing the previous paper here, but maybe we want to make it more emphasized in the Inro and across the paper that this is a follow-up of or previous paper and the setup is essentially the same. And cite the previous paper as "Paper 1" or BSM (sadly we don't have a second author with a surname starting with D...)}\rb{Should we invite Rebecca?}

\section{Results and Discussion}\label{sec:results}
Using the wind, circumstellar shell and SN parameters from Table~\ref{tab:ProgenitorModels}, we simulated the evolution of the remnants for 20 years for each of the 8 cases.
We evaluated the evolution of the particle distribution as well as the non-thermal gamma-ray and radio and the thermal X-ray and optical emission, using the methods outlined in sections \ref{sec:Absorption} and \ref{sec:Ap_rad}.

In \citetalias{Paper2} we discussed the evolution of the maximum energy, which can be significantly enhanced by the interaction with the shells in the LBV cases. Here, we will focus on the emission signatures that arise from the shell-interactions.

\subsection{Energy in comic rays}

We make no a-priori assumption on the explosion energy that gets converted into CRs and fix only the fraction of the thermal plasma's particles that get injected as CRs at the shock.  The total energy in CRs is an outcome of our simulation and depends on the the energy-flux through the shock or, alternatively, how the shock converts kinetic to thermal energy. One can easily find that the energy passing through the shock is given by
\begin{align}
   E_\text{acc} = \int \frac{1}{2}\rho_u v_\text{shock}^2 4\pi R_\text{shock}^2 v_\text{shock} \text{d}t \propto t^{3a-2} \label{eq:Eacc},
\end{align}
where $\rho_\mathrm{u}$ is the upstream density, $v_\text{shock}$ the shock velocity, $R_\text{shock}$ the shock radius and $a$ the expansion parameter. The point where $E_\text{acc}$ roughly equals the explosion energy marks the transition to the Sedov stage of the remnant's evolution. In the first few years, $E_\text{acc}$ is only small fraction of the $10^{51}\,$erg of explosion energy.

\begin{figure}[htb!]
    \centering
    \includegraphics[width=0.49\textwidth]{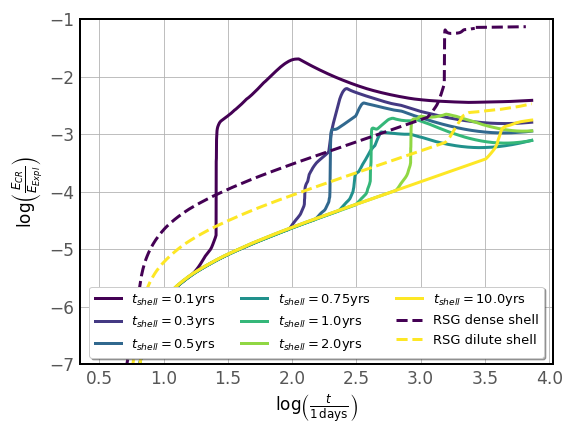}
    \caption{The total energy in CRs in units of $10^{51}\,$erg for the LBV-cases (solid) and RSG-cases (dashed) as a function of time.}
     \label{fig:Ecr}
\end{figure}

Figure \ref{fig:Ecr} shows the evolution of the total energy in CRs. Within the first day, the energy in CRs rises fast and then continues to increase roughly as $\propto t$, following from equation (\ref{eq:Eacc}) as our initial expansion parameter is $a\sim1$. However, the interaction with the shells breaks this power-law dependence, greatly increasing $E_\text{acc}$ and hence the energy in CRs.

The early-interaction LBV case reaches a peak energy-fraction of $\sim2\,$\%, marking the case with the highest conversion efficiency among the Type IIn cases. The RSG case interacting with the dense shell reaches $\sim 8\,$\% at the end of the simulation. Hence, the strong increase of $E_\text{max}$ and the gamma-ray luminosities (see section \ref{sec:gammarays}) is also related to the increased amount of energy that is available for CR acceleration at early times. After the shell interactions, the energy fraction decreases due-to adiabatic losses.

\subsection{Gamma-ray emission and detectability}
We calculated the gamma-ray emission as described in section \ref{sec:gamma-em} and applied the \gaga-absorption as described in section \ref{sec:gg_absorption}. First, we will discuss the emission in the 1-10\,TeV (hereafter \textit{VHE}) and 1-300\,GeV (hereafter \textit{Fermi}) energy bands. Additionally, we evaluated the gamma-ray detectability in the energy ranges for various current and future instruments which are shown only in visibility plots in section \ref{sec:gamma_vis} and appendix \ref{sec:Detectability}.

\subsubsection{Gamma-ray emission}\label{sec:gammarays}
We derived the $\gamma$-ray luminosity in the \textit{VHE} and \textit{Fermi} bands accounting only for hadronic emission. In this work, we consider a standard ISM composition for the CSM material, even though a heavier composition can be expected in the vicinity of massive stars. A heavier composition would not affect the light curves or luminosity considerably but could change the spectra shape of the gamma-ray emission, especially at energies below $1\,$GeV \citep{2020APh...12302490B, 2020MNRAS.497.3581D}.

%\is{IS: Robert, we talked about this, but just to put it here so we don't forget. Could be probably useful to calculate luminosities in the MeV energy range and compare to not approved but proposed missions like e-Astrogram (0.3 MeV - 3 GeV) and AMEGO-X (0.1 MeV - 1 GeV).}

\begin{figure}[h!]
    \centering
    \includegraphics[width=0.48\textwidth]{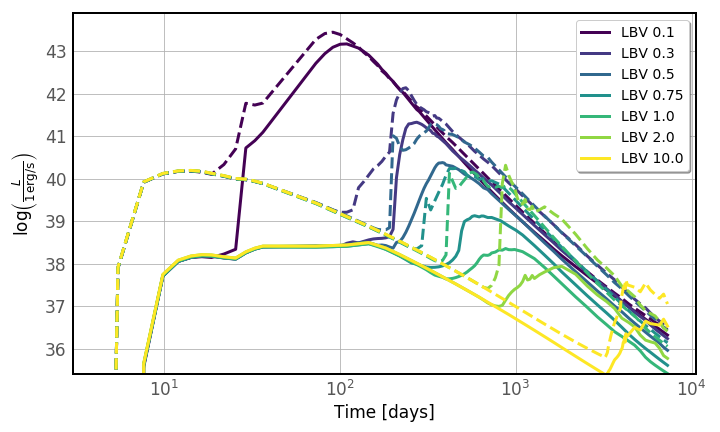}
    \includegraphics[width=0.48\textwidth]{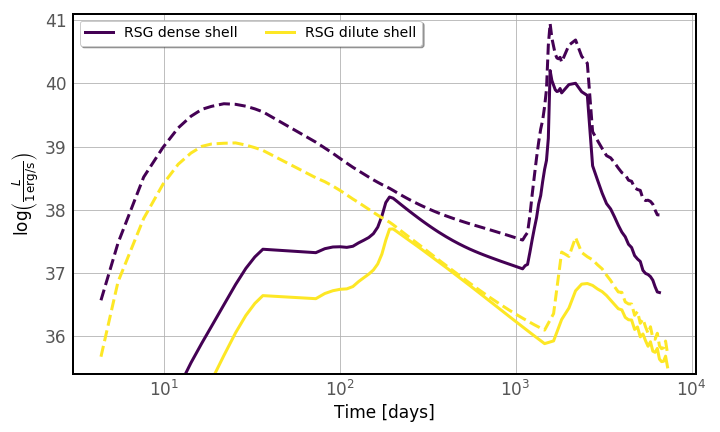}    
    \caption{Gamma-ray luminosities in the \textit{Fermi-LAT} energy range (dashed) and \textit{H.E.S.S.} energy range (solid) accounting for $\gamma\gamma$-absorption.\\
    \textbf{Top: }SNR of LBV progenitors interacting with the dense shell at different times (models as described in table \ref{tab:ProgenitorModels}). The $\gamma\gamma$-absorption has ceased completely after about 200 days. 
    \textbf{Bottom: }Interaction of SNRs with RSG progenitors interacting with dense shells created by external photoionization. The purple case represents an upper limit in terms of shell-mass, cause by a high mass-loss wind and an strong ambient photon field. The dilute case represents typical parameters for the population of RSGs. (See table~\ref{tab:ProgenitorModels} for details.)}
    \label{fig:GammaL}
\end{figure}

Figure \ref{fig:GammaL} illustrates the evolution of the $\gamma$-ray luminosity for our LBV and RSG models. The shock-shell interactions can clearly be seen as peaks in the $\gamma$-ray emission after the initial peak around tens of days that is caused by the interaction with the smooth wind. The peaks caused by the interactions  exceed the initial peak luminosities for LBV scenarios with interactions in the first two years, and for the RSG scenario with the massive shell. It is noteworthy that most IACTs undertake their pointed observations in the first weeks after the explosion, potentially missing late interactions. 

In fact, the $\gamma$-ray emission can, in the most extreme case of a LBV progenitor and an interaction after $\approx36\,$days, be five (three) orders of magnitude more luminous in the \textit{VHE} (\textit{Fermi}) band  compared to the early-time emission. This underlines that shock-shell interactions strongly enhance the detection-prospects of $\gamma$-ray emission from these objects. The interaction of the shock in the massive RSG-shell scenario still exceeds the initial luminosity by about one order of magnitude in \textit{Fermi}-band and two orders of magnitude in the \textit{VHE} band due to the stronger \gaga-absorption at higher energies. 

We reach gamma-ray luminosities as high as $10^{43}\,\text{erg s}^{-1}$ in the \textit{VHE} and Fermi-LAT bands for an early-interacting Type-IIn SNR and $10^{40}\,\text{erg s}^{-1}$ and $10^{41}\,\text{erg s}^{-1}$ for a Type-IIP SNR interacting with a dense shell in the \textit{VHE} and \textit{Fermi} bands respectively. These predictions are well above the current High-Energy Stereoscopic System (H.E.S.S.) upper limits of $\leq 1.5 \cdot 10^{39}\,\text{erg s}^{-1}$ for nearby Type-IIP events \citep{2019A&A...626A..57H}. However, these observations have been performed within the first year after explosion, where the flux in the \textit{VHE} band does not exceed $2\times10^{38}\,\text{erg s}^{-1}$ due to the strong \gaga-absorption and the insufficient densities. A close-by Type-IIn explosion would potentially have been detectable but no event occurred close enough to trigger H.E.S.S.~observations.

Interestingly, our heavy Type-IIP scenario produces a gamma-ray luminosity high enough to explain the detection of SN 2004dj in Fermi-LAT data about 5 years after the explosion \citep{2020ApJ...896L..33X}. Unfortunately, the SN occurred before the launch of Fermi-LAT but the onset of the observations coincides with the expected time for the shock-shell interaction in our models as does the $\approx5\,$year interval over which the signal fades. \cite{2020ApJ...896L..33X} report a luminosity of $\approx10^{39}\,\text{erg s}^{-1}$, right in in the middle of our predicted luminosity range: $3\times10^{37}-10^{41}\,\text{erg s}^{-1}$. 
Recently, \cite{2024A&A...686A.254M} observed the Type-IIP SNe SN 2023ixf located 6.85~Mpc from Earth and established a luminosity limit of $L_\gamma(E>1GeV)\leq6\cdot10^{40}-3.4\cdot10^{41}\,$erg. This limit is consistent with our prediction for the early-time emission in the freely expanding wind. After the shell interaction our predicted peak luminosity for the dense-shell case reaches the detection limit, if such a dense shell were present in the CSM of SN 2023ixf.

Our reference model from \citetalias{2022MNRAS.516..492B} used an unrealistically high mass-loss rate of $10^{-2}\,\mathrm{M}_\odot\,\mathrm{yr}^{-1}$. Even still, the peak luminosity of our shell-interaction model surpasses the luminosity of our previous LBV model by a factor of $\approx5$ in the \textit{VHE} band and reaches the same luminosity in the \textit{Fermi} band which is less affected by \gaga-absorption.

Due to the different absorption processes involved and due to the dynamic evolution of the particle-spectrum during the interaction, there are considerable time-shifts between the peaks of the emission in different energy bands. For our earliest Type-IIn interaction, there is first a peak in the HE gamma-ray emission around $\approx85\,$days, followed by a peak in the VHE domain after $\approx100\,$days and followed by a peak after around $150\,$days for the radio emission (see section \ref{sec:radio}). This poses an exceptional observational challenge in detecting these events as the HE-observatories operate in a survey mode and might only acquire enough photons after the VHE emission has peaked already. Usually the gamma-ray emission is fading already when the source peaks in radio. The time delay between HE and VHE gamma-ray emission can be larger for later interactions as the acceleration time for the highest-energy particles becomes larger.

\subsubsection{Gamma-ray detectability}\label{sec:gamma_vis}
%The higher peak-luminosities, that we reach mean, that the detection horizon for these objects is about a factor of 2 greater for instruments observing VHE $\gamma$-rays, and thus roughly $2\,Mpc$ and $6\,$Mpc for H.E.S.S. and CTAO-south respectively.
We evaluated the detectability of the emission for various present and future gamma-ray observatories. For compactness, the corresponding plots are presented in Appendix \ref{sec:Detectability}. %and tables of nearby Type-IIP and Type-IIn SNe in Appendix \ref{sec:SNR_list}.  %\jm{[Some sentences about how this is calculated would be good, or reference to methods section where it may be discussed?  e.g. for FERMI we continuously accumulate photons, but for IACTs we need to have dedicated pointings at specific times.  What do we assume about that?  e.g. assume we can trigger on X-ray/optical brightening and then observe for 50 hours over 10 nights?]}\rb{[I will add this...]}

We used the sensitivity from the Fermi-LAT collaboration\footnote{Obtained from https://fermi.gsfc.nasa.gov/ssc/data/analysis
/documentation/Cicerone/Cicerone\_LAT\_IRFs/LAT\_sensitivity.html} and from \cite{2016JPhCS.718e2043V} and \cite{cherenkov_telescope_array_observatory_2021_5499840} to estimate the detection horizon for survey-instruments Fermi-LAT, HAWC, LHAASO and the IACTs H.E.S.S.\footnote{Here taken as a representative of all current-generation IACTs.} and CTAO-South. We also included the projected eASTROGAM sensitivity from \citep{2017ExA....44...25D}. For the IACTs, we compared the flux at any given time during the remnant's evolution with the detection sensitivity for 50h of observations. For the survey-like instruments the required observation times can be longer than the duration of our transient signal. Therefore, we integrated the photon-flux since the explosion of the SN, denoted here as $t$, and compare the accumulated flux to the detection-sensitivity of the instruments scaled by $\sqrt{t_\text{ref}/t }$, where $t_\text{ref}$ denotes the observation time indicated for the sensitivity curves of Fermi-LAT, HAWC and their like. 

\textbf{Fermi-LAT} is %at the moment and in the foreseeable future 
the most sensitive gamma-ray telescope operating in the HE gamma-ray domain in survey mode, thus observing the whole sky continuously. It therefore should play an extraordinary role in detecting gamma-ray emission from nearby SNe and their remnants.
\begin{figure}[h!]
    \centering
    \includegraphics[width=0.45\textwidth]{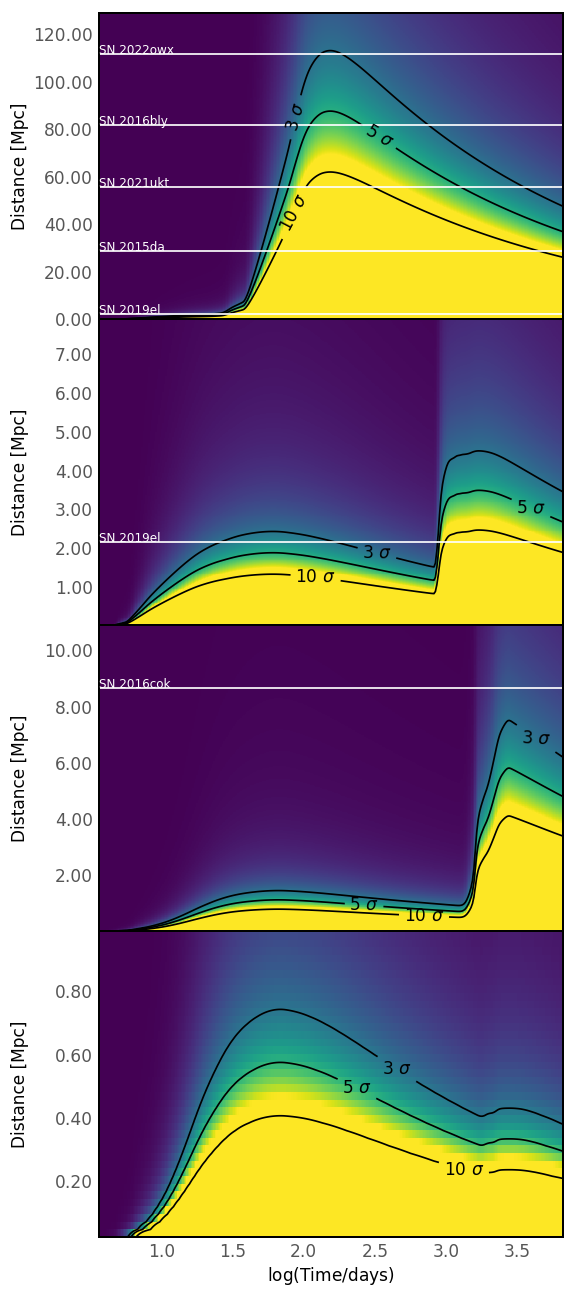}
    \caption{Fermi detectability for LBV 0.1, LBV 2.0, RSG high mass and RSG low mass scenarios from top to bottom. White lines indicate detected SNe explosions at their inferred distance and their overlap with operation-window of Fermi-LAT. \rb{Shown are Type-IIn events in the upper two panels and Type-IIP events in the lower panels.}}
    \label{fig:Fermi}
\end{figure}
Figure~\ref{fig:Fermi} shows the Fermi-LAT detectability of our two RSG scenarios and the LBV scenarios for interactions at 36~days and 2~years respectively. An early-interacting SNR would be detectable out to roughly $80\,$Mpc, although the significance rises only with the onset of the shell interaction after 36~days. The late interaction (after 2 years) would have a considerably smaller detection horizon of only 5~Mpc, again with the rise in significance occurring only with the onset of the interaction. The RSG scenario with the heavy shell enables a detection of up to 6~Mpc, while the low-mass shell interaction is only visible for SNe occurring within 600~kpc.\footnote{These numbers are somewhat higher than the values in \citetalias{2022MNRAS.516..492B} where, due to a bug in the production of the detectability plots, the denotes distances are too small by a factor of $\sim25$. An erratum is in preparation.} 

%There is a number of SNe that occurred sufficiently close-by to be potentially detectable (see tables \ref{tab:Type-IIn} and \ref{tab:Type-IIP}), if they were to have a dense and massive circumstellar shell like in one of our simulations. 
Since 2015 the Transient Name Server\footnote{\rb{The transient name server \citep[][https://www.wis-tns.org/]{2021AAS...23742305G} is the official channel to assign and classify transients by the IAU since 2015.}} lists a total of 39 firmly identified Type-IIn explosions that occurred at distances closer than 75~Mpc and one explosion - SN 2019el - at a distance closer than 6~Mpc. In the same time-frame only one Type-IIP event occurred at a distance closer than 10~Mpc - SN 2016cok - that is at the edge of detectability if a shell interaction occurs.
%Note that t
The recent nearby and bright type II SN 2023ixf and SN 2024ggi do not have a clear classification into either of these explosion types, %and so are not in these tables, 
although they are arguably the best candidates for gamma-ray detection on account of their proximity and evidence of circumstellar interaction.

The prospects at MeV energies using for instance the proposed eASTROGAM \citep{2017ExA....44...25D} are similar but limited to about half the distances of Fermi-LAT due to the pion-bump feature at lower energies (see also figure~\ref{fig:Astrogam}). 

Similarly to the increased detectability with Fermi-LAT, when shock-shell interactions take place the detection prospects increase also in the VHE domain using \textbf{H.E.S.S.} and other current-generation IACTs.
{Fig.~\ref{fig:hess} shows the VHE detectability of the same simulations as Fig.~\ref{fig:Fermi} using the sensitivity curve for H.E.S.S.\ as representative for the current generation of IACTs. 
The detection horizon increases to $\approx190(1.75)\,$Mpc for an early(/late) interacting Type-IIn explosion and $\approx10(0.6)\,$Mpc for the massive(/light) RSG shell. The detection prospects are best after $\approx100\,$days and 5~years for the LBV and RSG scenario, respectively. There is a shift between the peak detectability and onset of the interaction of about 100-1200 days in the LBV cases, as particles need time to get accelerated to energies where they radiate in the VHE domain. The peak in the HE domain is reached earlier, between 75-270 days after the onset of the interaction. An overview between the times of peaking in the different wavebands is shown in Table \ref{tab:Delay-times} and further discussed in Section \ref{sec:strategy}.
%\jm{[Is there a delay compared to FERMI? \rb{Yes, there is a delay !} This would be useful for HESS/CTAO if it was there.  Or is the time to accumulate enough Fermi photons longer than the delay time when the emission increase?  I guess we can see this from Fig.4: for 0.5-2-year delayed LBV interactions the Fermi peak is well before the TeV peak and could potentially trigger VHE observations.  For early and very late interactions the delay is not significant and we want to start VHE observations as soon as it becomes Gev-bright. This is important for planning an observing campaign if we can quantify (and explain) it. \rb{Quantifying is tricky as it depends on the exact flux-level. E.g., if its bright enough, you can trigger with Fermi but if its faint, you get a signal with fermi when its past the TeV-peak.}]}

The Transient Name Server lists a total of 94 SNe since 2015 that were close enough to be detected by H.E.S.S.~in case of an early interaction but none for a late interaction.
%\jm{[Important to quote here if any of these SNe actually had a re-brightening or other evidence for interaction with a really dense shell!  We could ask Takashi and Morgan Fraser, and maybe Nathan Smith?  Nathan would know every SN in the past 25 years that had a CSM shell interaction.  I really think the paper would benefit from input from some observers after this iteration.  These three could maybe help send the paper in more interesting directions that would get us cited by people outside the gamma-ray community.\rb{Sure :-). There was recently one report for a late-time radio brightening. But I have no overview how many cases are reported. Not all SNe are monitored well-enough, I guess...}]}
So far, observational indications of late-time interactions with CSM shells are rare and more late-time observations are  needed to get a reliable estimate of the number of potentially detectable SNe. A recent study by \cite{2025arXiv250201740S} found evidence for a brightening by at least two orders of magnitude in the radio band of the Type-IIb SNe SN 2001ig roughly 20 years after the explosion. 

Of the type IIP SNe since 2015, only SN 2016cok might be close enough for the detection of the RSG-shell interaction. The large potential number of SNe that could be detectable with H.E.S.S.~illustrates the need for a carefully crafted observation strategy. Without a trigger criterion for the onset of shock-shell interactions, monitoring of all potential SNRs is unfeasible. 

\cite{2019A&A...626A..57H} observed for instance SN 2008bk - a Type-IIP explosion at a distance of $4\pm0.4\,$Mpc - that could have been visible to H.E.S.S.~in a case of interaction with a circumstellar shell. However, given the distance and that H.E.S.S.~observations took place within the first year of explosion, the explosion would have needed to take place at least 4 times closer to Earth in order to be detectable. %\is{IS: should we maybe add this info in the D.1. table? To indicate for each SN whether it was observed by Fermi/HESS and how long after explosion? Only what is published of course. Or is it too much work?}\rb{Do we want to keep the table at all?}\is{Since you have it already, I think we could keep it. Would be helpful for others, I suppose. But maybe as an online material (not to count towards pages)}\rb{Hmm, due to the change in the hosting-erver of the transients, all the HESS-observed ones are actually not in our list...}

\textbf{HAWC and LHAASO} are both operating in survey mode but at slightly different energy ranges. In principle, SNe should be detectable to $60(0.5)\,$Mpc and $4(0.175)\,$Mpc for HAWC and $70(0.15)\,$Mpc and $6(0.2)\,$Mpc for LHAASO for the early(/late) LBV and heavy(/light) RSG scenarios respectively, as shown in Figs.~\ref{fig:hawc} and \ref{fig:lhaaso}, respectively.
Again, this is considerably farther than in the smooth-wind case \citepalias{2022MNRAS.516..492B}. In the case of LHAASO, the detectability horizon not only reflects the change in luminosity due to the different shell density but also the fact that later interactions lack the boost to $E_\text{max}$ and thus have a cutoff in the energy range where LHAASO is most sensitive. 
For both experiments, a total of 19 and 28 Type-IIn SNe have been close enough to potentially be detected for early interactions, while SN2016cok remains the only Type-IIP explosion that happened close enough for a potential detection. No explosion occurred close enough to be detectable in case of a late interaction or a light RSG shell.

\textbf{CTAO-South} is the only telescope still under construction that we consider here. The detectability distributions for CTAO-South look similar to the H.E.S.S.\ ones, but extend to much higher distances. In fact, by \rb{reaching $300(3)\,$Mpc }for the Type-IIn case, CTAO-South will probe distances where EBL absorption becomes a factor. Type-IIP explosions still remain detectable only in the vicinity up \rb{to $\approx20(1.2)\,$Mpc}. 
While CTAO-south would potentially have been able to follow-up on 94 Type-IIn explosions, similar to the existing IACTs, there are \rb{6} additional Type-IIP events that are below the optimistic detection threshold.

\subsection{Radio emission}\label{sec:radio}
We calculated the radio emission based on the electron distribution and the magnetic field, including the self-amplified component using the method as described in \citepalias[][]{2022MNRAS.516..492B}. Figures \ref{fig:RadioIndex1} and \ref{fig:RadioIndex2} show the radio luminosity at $4\,$GHz including the effects of free-free absorption in the CSM, and the evolution of the radio spectral index, for the LBV and RSG shell-interaction simulations, respectively.
Where \citetalias{2022MNRAS.516..492B} struggled to explain the observed peak luminosity of the radio emission either by strong free-free absorption or too low magnetic fields for moderate mass-loss, the shock-shell interaction overcomes both shortcomings in the LBV scenarios. As soon as the shock has passed the majority of the dense shell, the radio emission can freely escape whereas the interaction itself boosts magnetic field amplification to levels commensurate with the observed radio-flux. The observed peak is now in good agreement with the population study of radio SNe from \cite{2021ApJ...908...75B}. We also added the observed radio lightcurves for a number of SNe in both plots. While the onset of our predicted emission is in rough agreement with the measurements, our predicted luminosities seem to be slightly too low, however, the most well-charcterized SNe in this sample from \cite{2021ApJ...908...75B} tend also to be the brightest ones. We also added the lightcurve for SN 2001ig, where a sign of radio-rebrightening was recently detected \citep{2025arXiv250201740S}. Besides SN 2001ig being classified as a Type-IIb event, the observed luminosity of the rebrightening matches our prediction for a shell-interaction well.

%We calculated the radio emission based on the electron distribution and the magnetic field, including the self-amplified component. Figure \ref{fig:RadioIndex} shows the radio luminosity at $8\,$GHz, including the effects of free-free absorption in the CSM, and the evolution of the radio spectral index.
\begin{figure*}
   \centering
    \includegraphics[width=0.95\textwidth]{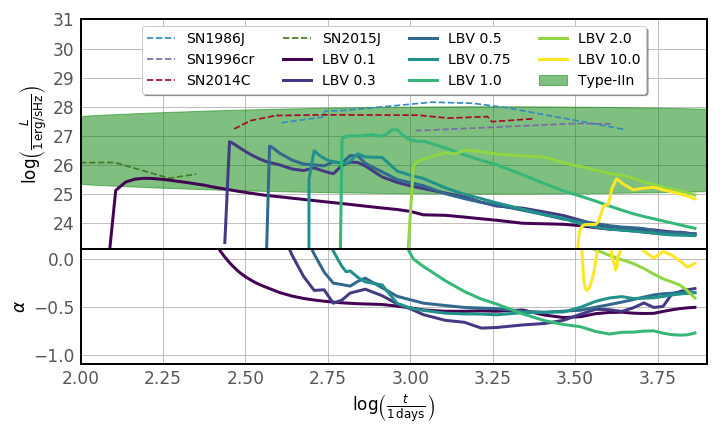}
    \caption{\textbf{Top panel:} Radio luminosity including the effects of free-free absorbtion for Type-IIn explosions. The green area indicates the $1\sigma$ uncertainty region for the rise time and peak radio-luminosity for Type-IIn SN \rb{taken from \cite{2021ApJ...908...75B}}. \textbf{Bottom panel:} Radio spectral index $\alpha$ of the absorbed radio flux at 4\,GHz.}
    \label{fig:RadioIndex1}
\end{figure*}
\begin{figure*}
   \centering    
    \includegraphics[width=0.95\textwidth]{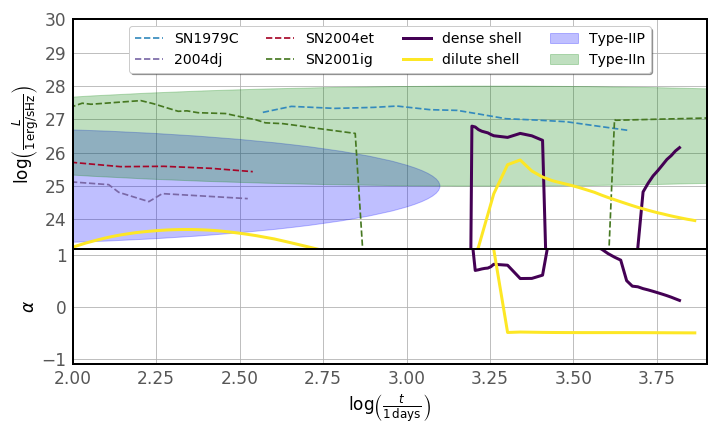}
    \caption{Same as Figure~\ref{fig:RadioIndex1} but for the RSG/Type-IIP scenarios. \rb{The Type-IIP population average is taken from \cite{2021ApJ...908...75B}}}
    \label{fig:RadioIndex2}
\end{figure*}

Interestingly, the shell interaction affects the radio spectral index by softening it. When absorption becomes negligible, the radio index is softer than $\alpha=-0.5$ and reaching values as low as $\alpha=-0.75$ in the Type-IIn scenarios but gradually hardens over time. The reason is that during the shell interaction at each succeeding time, more particles are injected than before. As the acceleration is not instantaneous, this causes a softening of the spectrum. The effect fades once the shock has passed the shell and, when recently accelerated particles start to dominate, the spectrum hardens. A similar effect of a gradual radio-hardening, starting from a soft spectrum, can be seen in the case of SN1987A \citep{2010ApJ...710.1515Z}. There, the shock interacted with a dense equatorial ring that sparked a strong brightening in the radio and X-ray emission of the remnant - a situation that is, despite the different geometry, comparable to our simulation setup.

For the Type-IIP scenarios, we see a double peak structure, where a first peak emerges after 100s of days in the low-mass case, followed by a second peak when the shock-shell interaction happens. The spectrum is dominated by a non-thermal spectrum once the shock passed most of the shell and the spectral-index reaches $\alpha\approx-0.5$ rather quickly.
In the high-mass case, the first peak is suppressed due to the stronger absorption in the more dense shell. The shock-shell interaction boosts the radio emission, however, the spectrum appears to be thermal as the high shell mass compared to the ejecta mass slows down the shock considerably, preventing the formation of a non-thermal spectrum in that case. Only after the reacceleration of the shock, the non-thermal radio emission starts to rise $\approx$15~years post-explosion.}

Observations suggest radio-spectra even softer than $\alpha\approx-0.75$ for Type-Ib and Type-Ic supernovae, indicating electron-spectra as soft as $s=-3$ \citep{2006ApJ...651..381C}. Type-IIP explosions on the other hand have been modeled with electron-spectra closer to $s=-2.2$. \cite{2006ApJ...641.1029C} pointed out that cooling can significantly soften the spectra in these cases.

\subsection{Thermal X-ray emission}
As a proxy for the thermal X-ray emission, we follow the ansatz described in sections \ref{sec:Xray} and \ref{sec:xray_abs}. In both progenitor cases, the interaction with the shell strongly boosts the thermal X-ray emission, by up to five orders of magnitude for the LBVs and four orders of magnitude for the RSGs. However, the boost in the LBV case has to be considered as an upper limit as the shock is radiative for the high densities present during the shell interaction.

\begin{figure*}[hbt!]
   \centering
    \includegraphics[width=0.99\textwidth]{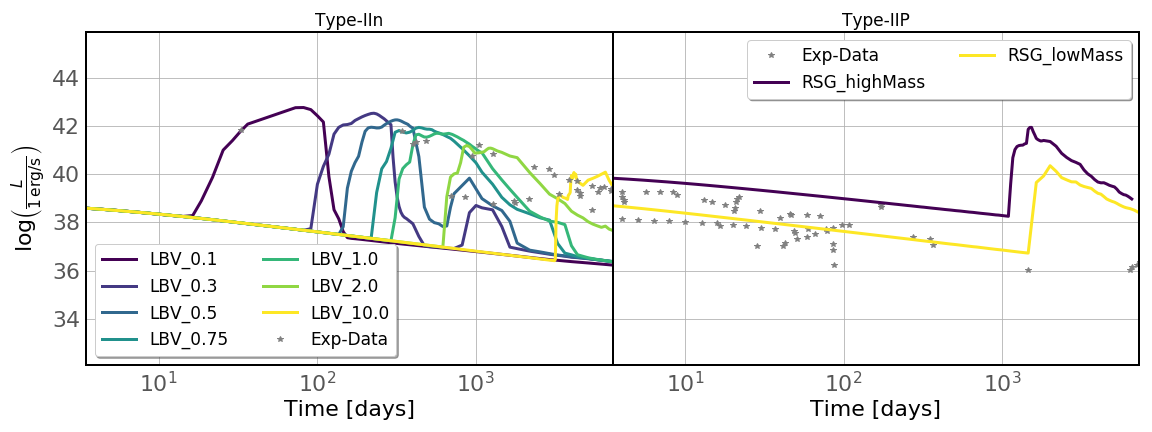}
    \caption{Absorbed thermal X-ray luminosity between 0.2~keV to 10~keV for a Type-IIn explosions on the left and Type-IIP explosions on the right. The grey stars indicate measurements listed in  \citet{2014MNRAS.440.1917D}.}
    \label{fig:ThermalXray}
\end{figure*}

Figure \ref{fig:ThermalXray} shows that our predicted X-ray emission is roughly consistent with the observational constraints. However, there is no experimental detection of X-rays from LBVs before approximately one year post-explosion, which could simply be an observational bias or point to typically larger distances of the shells from the SN progenitor.
 
For Type-IIP explosions, our early emission basically agrees with the observations by design, as we use typical mass-loss rates. The shell interaction then greatly enhances this emission. However, at the onset of the interaction after a few years, targeted X-ray observations of previously detected SNe are rare, as no coordinated long-term monitoring program exists. There is a strong probability that these rebrightening events are missed observationally. 
Only an X-ray survey instrument can realistically hope to detect these shell-interaction events without a trigger from another waveband.

\subsection{Optical emission}
The difficulty of detecting shock-shell interactions, especially for RSG progenitors, is the large time-offset between the interaction and the explosion. Here, targeted observations as carried out in the radio or X-ray band are rare this late after the explosions and hence many potentially interesting interactions might be missed. At the moment, a search for flaring transients with Fermi-LAT might be a way to provide triggers but a sufficiently large observation window is required. Further, future IACTs have a notably farther detection horizon, so that Fermi-LAT would not be sensitive enough to provide triggers for e.g. CTAO.

\begin{figure}[hbt!]
   \centering
    \includegraphics[width=0.49\textwidth]{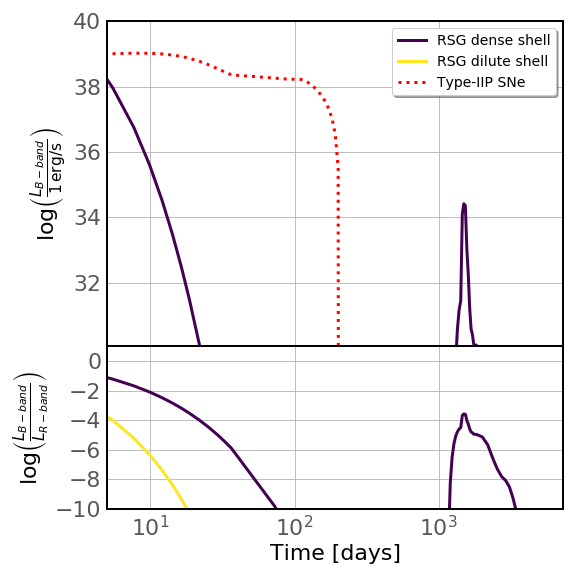}
    \caption{\textbf{Top: } Optical lightcurves in the B-band (398nm-492nm) for SN interaction with the two RSG shells (solid lines) and the luminosity from the SN photosphere (dotted red line).
    \textbf{Bottom: } Ratio of the B-band and R-band  (591nm-727nm) luminosities over time.}
    \label{fig:Optical}
\end{figure}

An alternative are the currently operating high-cadence optical surveys as ZTF \citep{2019PASP..131a8002B}, Pan-STARRS \citep{2004AN....325..636H} or Gaia \citep{2013RSPTA.37120239H}, that regularly monitor large parts of the night sky. Following the ansatz described in section \ref{sec:Optical_method}, we calculated the optical emission expected from the interaction for the RSG cases. 
Figure \ref{fig:Optical} shows the optical luminosity in the B-band and the ratio of the B-band and R-band luminosities for the RSG case. The shell interaction gives rise to enhanced X-ray emission and likewise to optical emission that is -- depending on the shell mass -- 5 orders of magnitude below the peak of the emission from the SN photosphere. This means that the most dense SNR-shell interactions for RSGs might be detectable out to a few Mpc, given the detection horizon of Type-IIP SNe of up to a few hundred Mpc. However, less dense shells are considerably weaker in their optical emission and hence harder to detect.

\section{Observation strategy}\label{sec:strategy}
\begin{table*}[h]
  \centering
  \caption{Time of first interaction with shell and time till peak in corresponding waveband after the onset of the interaction.}
  \label{tab:Delay-times}
  \begin{tabular}{l | c c c c c}
Model &		Interaction time [yr]& 	$\Delta t_\text{Fermi}$[d]& 	$\Delta t_\text{Hess}$[d]& 	$\Delta t_\text{Xray}$[d]& 	$\Delta t_\text{Radio}$[d]\\
\hline
LBV 0.1& 	0.04&	76& 	    94& 	    67& 	    140\\
LBV 0.3& 	0.236& 	151& 	187& 	141& 	196\\
LBV 0.5& 	0.377& 	200& 	245& 	181& 	236\\
LBV 0.75& 	0.594& 	239& 	367& 	221& 	294\\
LBV 1.0& 	0.809& 	270& 	580& 	270& 	617\\
LBV 2.0& 	1.71& 	251& 	1200& 	251& 	835\\
LBV 10.0& 	8.65& 	1076& 	3777& 	3412& 	1076\\
RSG highMass& 	3.0& 	474& 	474& 	438& 	474\\
RSG lowMass& 	4.13& 	682& 	1047& 	500& 	682\\
  \end{tabular}
\end{table*}

Our work suggests that current observation strategies for detecting gamma-ray emission from CC-SNe could be improved. %might be not promising. 
The initial \gaga-absorption attenuates the signal, while the rise in the light curves associated with the potential shock-shell interaction can happen rather late \citep{2015ApJ...815..120M}. Hence, a fundamental problem in detecting the gamma-ray emission from CC-SNe in our scenarios is defining a suitable trigger for targeted observations. 

Table \ref{tab:Delay-times} indicates that the X-ray emission peaks first in practically all cases, as it directly reflects the change in the ISM density. However, Fermi-LAT, where the peak luminosity is reached tens of days after the peak on thermal X-rays, is the only survey instrument in the HE domain that is currently operational and needs, depending on the source flux, integration times of tens to hundreds of days for a detection. The emission at VHE-energies peaks in general after the peak in HE-energies is reached. As a consequence, the peak of the emission as detectable with IACTs might be missed as Fermi-LAT needs a considerable amount of integration time to detect a signal. The radio emission in general peaks last due to the absorption in the dense shell. On the other hand, radio observations have the potential to monitor nearby SNe for signals of rebrightening as done for SN 2001ig \citep{2025arXiv250201740S}. \rb{We note though that the timing in Table \ref{tab:Delay-times} is idealized in our spherically symmetric treatment and might be altered, especially smeared, if the explosions itself or the CSM structure significantly differs from spherical symmetry.}

Emission in the optical might also provide good prospects for a detection, as large parts of the sky can be monitored with a high cadence. As for a potential detection in radio, a timely communication to the high-energy community would be crucial in order to have a chance of a detection in gamma-rays.

Furthermore, for our Type-IIn models, it is evident that the peak of the gamma-ray emission coincides more closely with the peak of the radio emission than with the optical emission of the SN explosion. The shell interaction not only increases the target density for pion production, but the higher ambient density means that more particles -- including electrons -- can be accelerated and, consequently, a brightening in radio and gamma-rays has a common origin. Again, the position of the (first) peak in the radio emission, can be hundreds to thousands of days after the explosion: considerably later than the current timescales for follow-up observations.

Possible follow-up instruments in the centimeter and millimeter radio bands are MeerKAT \citep{jonas16} and the Atacama Large Millimeter Array \cite[ALMA,][]{wootten09}, both the most sensitive existing instruments in their wavelength regime. Considering sensitivities of 50~$\mu$Jy~bm$^{-1}$ at 100~GHz, and 15~$\mu$Jy~bm$^{-1}$ at 3~GHz (for a 15 minute ALMA integration and a 12 minute MeerKAT integration, respectively) our models predict that the emission would always be detectable first in the ALMA band. For the cases considered, for the LBVs, the shortest timescale at which the source would be detectable occurs for the LBV 10.0 case (as early as one week post-explosion for a 1~Mpc distance), and the longest for the LBV 2.0 case (where emission is not detectable until two years post-explosion, also for a 1~Mpc distance). For the RSG cases the emission should be detectable with ALMA in the first five days post-explosion (for the low-mass case) and in the first three weeks (for the high-mass case), again for a 1~Mpc distance. How late after an ALMA detection would the emission become detectable with MeerKAT varies with a timescale of days to years for the cases considered.

For this reason, radio observations can provide important diagnostics that could be used as triggers for deep gamma-ray observations. In several cases the ALMA detections happen well ahead of the shell interaction, and can track the brightening associated with it. The lower radio frequency observations probe the absorption or lack thereof due to the presence of a shells of different masses and distances to the star. When combined, it is possible to discriminate between cases, and constrain the scenario a given event represents, and hence the most optimal time to conduct the costly gamma-ray observations. Such a strategy of monitoring the radio luminosity with \textit{ALMA} or \textit{MeerKAT} is likely feasible for the few supernovae that occur within $\lesssim 25$~Mpc.

Apart from a dedicated ALMA target of opportunity project as suggested above, the gamma-ray community would benefit from closer and more timely communication with the many teams that follow nearby supernovae with a wide variety of radio facilities such as the Karl G.~Jansky Very Large Array (JVLA), the Giant Metre Wave Radio Telescope (GMRT), or the Australia Telescope Compact Array (ATCA), often for years post-explosion. 

It is worth noting that the cases discussed here are physically distinct from observed SN such as SN 2024ggi, where a dense CSM was detected up to roughly $2.3\times10^{-14}$cm around the progenitor with a mass-loss rate of roughly $\dot{M}=10^{-2}M_\odot$\,yr$^{-1}$ and $v_\text{wind}=40\,$km\,s$^{-1}$ \citep{2024ApJ...972L..15S}.
Similarly, a dense CSM extending to $r\sim10^{15}$\,cm was inferred from modelling of SN 2023ixf \citep[e.g.][]{2025A&A...694A.319K}.
These values roughly correspond to the density setup for the free wind in our LBV scenarios. However, it is evident from our models that such an event would only be detectable to distances of up to $4\,$Mpc depending on the instrument, while SN 2024ggi happened at a distance of roughly $7\,$Mpc. Further, the small extension of the dense CSM in that case means that a high internal gamma-ray luminosity is only achieved within the first few days post-explosion, where \gaga-absorption is strong and efficiently attenuating the signal \citepalias[see also][]{2022MNRAS.516..492B}. Cases like SN 2024ggi are likely linked to enhanced RSG mass-loss just prior to their explosion \citep{2021ApJ...912...46B} but is not creating the additional shell-like structures, that we discussed here.

%\begin{itemize}
%\item Offset between wavelengths: HE gamma-rays, VHE gamma-rays, (X-ray(?), Radio
%\item current approach: early on
%\item Trigger needed: monitoring of SNe for re-brightening in X-ray/radio (optical?)
%\item TRIGGER ON RADIO SNR/PEAK instead of very early on
%\item Type-IIP interesting targets: Shell confirmed for Betelgeuse; SNe with hint; abundant
%\item Type-IIP: consistent with radio
%\item Type-IIn rare but very interesting (PeVatrons)
%\item Connection to FBOTs?
%\end{itemize}

\section{Conclusions}\label{sec:conclusions}
We performed numerical simulations of particle acceleration in very young SNRs expanding in dense circumstellar media featuring dense shells created by the progenitors, solving time-dependent transport equations of CRs and magnetic turbulence in the test-particle limit alongside the standard gas-dynamical equations for CC-SNRs. 

The peak luminosity in the $\gamma$-ray domain reached during the shock-shell interaction is well exceeding the initial luminosities of the first weeks after the explosion (that are strongly affected by \gaga-absorption). In the VHE domain, the peak luminosity during the interaction is about a factor of 5 higher then the luminosity expected from a progenitor with a smooth wind that is about a factor of 100 denser then the case that we consider. We obtain luminosties of up to $10^{43}\,$erg $s^{-1}$ for early-interacting Type-IIn explosions and $10^{40}\,$erg $s^{-1}$ and $10^{41}\,$erg $s^{-1}$ for the light and heavy shell models of RSG progenitors. 

We investigated the radio emission, taking into account the effect of free-free absorption in the ambient medium. We find peak luminosities consistent with the population average for Type-IIn explosions. The radio spectral index appears softer than $\alpha=-0.5$ right after the free-free absorption became negligible and the emission detectable, and gradually hardens towards the canonical $\alpha=-0.5$, similar to the evolution observed for SN 1987A.

The late peak of the gamma-ray emission is at odds with current observation strategies of IACTs. After evaluating the thermal emission in the optical and x-rays, which both peak after the gamma-ray emission, we identify high-cadence optical surveys as a potentially suitable tool to capture the most extreme Type-IIPs interacting with dense shells. Due to the low sample sizes in the nearby universe and the short required observation times, a systematic radio and mm monitoring of close-by SNe for a few years after their explosions might prove valuable for identifying late shell-interactions as well.

Current generation instruments can detect LBV shell interaction up to 80~Mpc and 190~Mpc for the most extreme cases, while RSG shell interactions are only visible up to 6~Mpc and 10~Mpc for Fermi-LAT and H.E.S.S. respectively. CTAO could push these boundaries \rb{to 300~Mpc and 20~Mpc} for our most extreme LBV and RSG scenarios.

\section*{Data availability}
The data underlying this article will be shared on reasonable request to the corresponding author.
\section*{Acknowledgements}
RB and JM acknowledge funding from an Irish Research Council Starting Laureate Award. 
JM acknowledges funding from a Royal Society-Science Foundation Ireland University Research Fellowship.
This publication results from research conducted with the financial support of Taighde \'Eireann - Research Ireland under Grant numbers 20/RS-URF-R/3712, IRCLA\textbackslash 2017\textbackslash 83.
We acknowledge the SFI/HEA Irish Centre for High-End Computing (ICHEC) for the provision of computational facilities and support (project dsast026c). IS acknowledges funding from Comunidad de Madrid through the Atracción de Talento “César Nombela” grant with reference number 2023-T1/TEC-29126.

We are grateful to Morgan Fraser and Takashi Moriya for their valuable input and discussions on the paper.
%%%%%%%%%%%%%%%%%%%%%%%%%%%%%%%%%%%%%%%

\bibliographystyle{aa}
\bibliography{References}

\begin{thebibliography}{86}
\expandafter\ifx\csname natexlab\endcsname\relax\def\natexlab#1{#1}\fi

\bibitem[{Abdalla {et~al.}(2021)Abdalla, Aharonian, Ait-Benkhali, Anguener, Arcaro, Armand, Armstrong, Ashkar, Backes, Baghmanyan, Barbosa~Martins, Barnacka, Barnard, Batzofin, Becherini, Berge, Bernloehr, Bi, Böttcher, Boisson, Bolmont, Bony~(de), Breuhaus, Brose, Brun, Bulik, Bylund, Cangemi, Caroff, Casanova, Catalano, Chambery, Chand, Chen, Cotter, Curlo, Dalgleish, Damascene~Mbarubucyeye, Davids, Davies, Devin, Djannati-Ataï, Dmytriiev, Donath, Doroshenko, Dreyer, Du~Plessis, Duffy, Egberts, Einecke, ERNENWEIN, Fegan, Feijen, Fiasson, Fichet~de Clairfontaine, Fontaine, Frans, Fuessling, Funk, Gabici, Gallant, Giavitto, Giunti, Glawion, Glicenstein, Grondin, Hattingh, Haupt, HERMANN, Hinton, Hofmann, Hoischen, Holch, Holler, Horns, Huang, Huber, Hörbe, Jamrozy, Jankowsky, Joshi, JUNG, Kasai, Katarzynski, Katz, Khangulyan, Khelifi, Klepser, Kluzniak, Komin, Konno, Kosack, Kostunin, Kreter, Kukec~Mezek, Kundu, Lamanna, Le~Stum, Lemiere, Lemoine-Goumard, Lenain, Leuschner, Levy, Lohse, Luashvili, Lypova,
  Mackey, Majumdar, Malyshev, MALYSHEV, Marandon, Marchegiani, Marcowith, Mares, Marti'i-Devesa, Marx, Maurin, Meintjes, Meyer, Mitchell, Moderski, Mohrmann, Montanari, Moore, Morris, Moulin, Muller, Murach, Nakashima, Naurois~(de), Nayerhoda, Davids, Niemiec, Noel, O'Brien, Oberholzer, Ohm, Olivera-Nieto, Ona-Wilhelmi~(de), Ostrowski, Panny, Panter, Parsons, Peron, Pita, Poireau, Prokhorov, Prokoph, PUEHLHOFER, Punch, Quirrenbach, Reichherzer, Reimer, Reimer, Remy, Renaud, Reville, Rieger, Romoli, Rowell, Rudak, Rueda~Ricarte, Ruiz~Velasco, Sahakian, Sailer, Salzmann, Sanchez, Santangelo, Sasaki, Schaefer, Schutte, Schwanke, Schüssler, Senniappan, Seyffert, Shapopi, Shiningayamwe, Simoni, Sinha, Sol, Spackman, Specovius, Spencer, Spir-Jacob, Stawarz, Steenkamp, Stegmann, Steinmassl, Steppa, Sun, Takahashi, Tanaka, Tavernier, Taylor, Terrier, Thiersen, Thorpe-Morgan, Tluczykont, Tomankova, Tsirou, Tsuji, Tuffs, Uchiyama, van~der Walt, van Eldik, van Rensburg, van Soelen, Vasileiadis, Veh, Venter, Vincent,
  Vink, Völk, Wagner, Watson, Werner, White, Wierzcholska, Wong, Yassin, Yusafzai, Zacharias, Zanin, Zargaryan, Zdziarski, Zech, Zhu, Zmija, Zouari, Żywucka, \& Ryder}]{Abdalla:2021Mg}
Abdalla, H., Aharonian, F., Ait-Benkhali, F., {et~al.} 2021, in Proceedings of 37th International Cosmic Ray Conference {\textemdash} PoS(ICRC2021), Vol. 395, 809

\bibitem[{{Abeysekara} {et~al.}(2020){Abeysekara}, {Archer}, {Benbow}, {Bird}, {Brose}, {Buchovecky}, {Buckley}, {Chromey}, {Cui}, {Daniel}, {Das}, {Dwarkadas}, {Falcone}, {Feng}, {Finley}, {Fortson}, {Gent}, {Gillanders}, {Giuri}, {Gueta}, {Hanna}, {Hassan}, {Hervet}, {Holder}, {Hughes}, {Humensky}, {Kaaret}, {Kar}, {Kelley-Hoskins}, {Kertzman}, {Kieda}, {Krause}, {Krennrich}, {Kumar}, {Lang}, {Maier}, {Moriarty}, {Mukherjee}, {Nievas-Rosillo}, {O'Brien}, {Ong}, {Park}, {Petrashyk}, {Pfrang}, {Pohl}, {Pueschel}, {Quinn}, {Ragan}, {Reynolds}, {Richards}, {Roache}, {Sadeh}, {Santander}, {Sembroski}, {Shahinyan}, {Sushch}, {Weinstein}, {Wilcox}, {Wilhelm}, {Williams}, {Williamson}, {Zitzer}, \& {Ghiotto}}]{2020ApJ...894...51A}
{Abeysekara}, A.~U., {Archer}, A., {Benbow}, W., {et~al.} 2020, \apj, 894, 51

\bibitem[{{Acciari} {et~al.}(2022){Acciari}, {Ansoldi}, {Antonelli}, {Arbet Engels}, {Artero}, {Asano}, {Baack}, {Babi{\'c}}, {Baquero}, {Barres de Almeida}, {Barrio}, {Batkovi{\'c}}, {Becerra Gonz{\'a}lez}, {Bednarek}, {Bellizzi}, {Bernardini}, {Bernardos}, {Berti}, {Besenrieder}, {Bhattacharyya}, {Bigongiari}, {Biland}, {Blanch}, {B{\"o}kenkamp}, {Bonnoli}, {Bo{\v{s}}njak}, {Busetto}, {Carosi}, {Ceribella}, {Cerruti}, {Chai}, {Chilingarian}, {Cikota}, {Colak}, {Colombo}, {Contreras}, {Cortina}, {Covino}, {D'Amico}, {D'Elia}, {Da Vela}, {Dazzi}, {De Angelis}, {De Lotto}, {Del Popolo}, {Delfino}, {Delgado}, {Delgado Mendez}, {Depaoli}, {Di Pierro}, {Di Venere}, {Do Souto Espi{\~n}eira}, {Prester}, {Donini}, {Dorner}, {Doro}, {Elsaesser}, {Fallah Ramazani}, {Fari{\~n}a Alonso}, {Fattorini}, {Fonseca}, {Font}, {Fruck}, {Fukami}, {Fukazawa}, {Garc{\'\i}a L{\'o}pez}, {Garczarczyk}, {Gasparyan}, {Gaug}, {Giglietto}, {Giordano}, {Gliwny}, {Godinovi{\'c}}, {Green}, {Green}, {Hadasch}, {Hahn}, {Hassan}, {Heckmann},
  {Herrera}, {Hoang}, {Hrupec}, {H{\"u}tten}, {Inada}, {Ishio}, {Iwamura}, {Jim{\'e}nez Mart{\'\i}nez}, {Jormanainen}, {Jouvin}, {Kerszberg}, {Kobayashi}, {Kubo}, {Kushida}, {Lamastra}, {Lelas}, {Leone}, {Lindfors}, {Linhoff}, {Lombardi}, {Longo}, {L{\'o}pez-Coto}, {L{\'o}pez-Moya}, {L{\'o}pez-Oramas}, {Loporchio}, {Machado de Oliveira Fraga}, {Maggio}, {Majumdar}, {Makariev}, {Mallamaci}, {Maneva}, {Manganaro}, {Mannheim}, {Maraschi}, {Mariotti}, {Mart{\'\i}nez}, {Mas Aguilar}, {Mazin}, {Menchiari}, {Mender}, {Mi{\'c}anovi{\'c}}, {Miceli}, {Miener}, {Miranda}, {Mirzoyan}, {Molina}, {Moralejo}, {Morcuende}, {Moreno}, {Moretti}, {Nakamori}, {Nava}, {Neustroev}, {Nievas Rosillo}, {Nigro}, {Nilsson}, {Nishijima}, {Noda}, {Nozaki}, {Ohtani}, {Oka}, {Otero-Santos}, {Paiano}, {Palatiello}, {Paneque}, {Paoletti}, {Paredes}, {Pavleti{\'c}}, {Pe{\~n}il}, {Persic}, {Pihet}, {Prada Moroni}, {Prandini}, {Priyadarshi}, {Puljak}, {Rhode}, {Rib{\'o}}, {Rico}, {Righi}, {Rugliancich}, {Sahakyan}, {Saito}, {Sakurai},
  {Satalecka}, {Saturni}, {Schleicher}, {Schmidt}, {Schweizer}, {Sitarek}, {{\v{S}}nidari{\'c}}, {Sobczynska}, {Spolon}, {Stamerra}, {Stri{\v{s}}kovi{\'c}}, {Strom}, {Strzys}, {Suda}, {Suri{\'c}}, {Takahashi}, {Takeishi}, {Tavecchio}, {Temnikov}, {Terzi{\'c}}, {Teshima}, {Tosti}, {Truzzi}, {Tutone}, {Ubach}, {van Scherpenberg}, {Vanzo}, {Vazquez Acosta}, {Ventura}, {Verguilov}, {Vigorito}, {Vitale}, {Vovk}, {Will}, {Wunderlich}, {Yamamoto}, {Zari{\'c}}, \& {Ambrosino}}]{2022NatAs...6..689A}
{Acciari}, V.~A., {Ansoldi}, S., {Antonelli}, L.~A., {et~al.} 2022, Nature Astronomy, 6, 689

\bibitem[{{Ackermann} {et~al.}(2014){Ackermann}, {Ajello}, {Albert}, {Baldini}, {Ballet}, {Barbiellini}, {Bastieri}, {Bellazzini}, {Bissaldi}, {Blandford}, {Bloom}, {Bottacini}, {Brandt}, {Bregeon}, {Bruel}, {Buehler}, {Buson}, {Caliandro}, {Cameron}, {Caragiulo}, {Caraveo}, {Cavazzuti}, {Charles}, {Chekhtman}, {Cheung}, {Chiang}, {Chiaro}, {Ciprini}, {Claus}, {Cohen-Tanugi}, {Conrad}, {Corbel}, {D'Ammando}, {de Angelis}, {den Hartog}, {de Palma}, {Dermer}, {Desiante}, {Digel}, {Di Venere}, {do Couto e Silva}, {Donato}, {Drell}, {Drlica-Wagner}, {Favuzzi}, {Ferrara}, {Focke}, {Franckowiak}, {Fuhrmann}, {Fukazawa}, {Fusco}, {Gargano}, {Gasparrini}, {Germani}, {Giglietto}, {Giordano}, {Giroletti}, {Glanzman}, {Godfrey}, {Grenier}, {Grove}, {Guiriec}, {Hadasch}, {Harding}, {Hayashida}, {Hays}, {Hewitt}, {Hill}, {Hou}, {Jean}, {Jogler}, {J{\'o}hannesson}, {Johnson}, {Johnson}, {Kerr}, {Kn{\"o}dlseder}, {Kuss}, {Larsson}, {Latronico}, {Lemoine-Goumard}, {Longo}, {Loparco}, {Lott}, {Lovellette}, {Lubrano},
  {Manfreda}, {Martin}, {Massaro}, {Mayer}, {Mazziotta}, {McEnery}, {Michelson}, {Mitthumsiri}, {Mizuno}, {Monzani}, {Morselli}, {Moskalenko}, {Murgia}, {Nemmen}, {Nuss}, {Ohsugi}, {Omodei}, {Orienti}, {Orlando}, {Ormes}, {Paneque}, {Panetta}, {Perkins}, {Pesce-Rollins}, {Piron}, {Pivato}, {Porter}, {Rain{\`o}}, {Rando}, {Razzano}, {Razzaque}, {Reimer}, {Reimer}, {Reposeur}, {Saz Parkinson}, {Schaal}, {Schulz}, {Sgr{\`o}}, {Siskind}, {Spandre}, {Spinelli}, {Stawarz}, {Suson}, {Takahashi}, {Tanaka}, {Thayer}, {Thayer}, {Thompson}, {Tibaldo}, {Tinivella}, {Torres}, {Tosti}, {Troja}, {Uchiyama}, {Vianello}, {Winer}, {Wolff}, {Wood}, {Wood}, {Wood}, {Charbonnel}, {Corbet}, {De Gennaro Aquino}, {Edlin}, {Mason}, {Schwarz}, {Shore}, {Starrfield}, {Teyssier}, \& {Fermi-LAT Collaboration}}]{2014Sci...345..554A}
{Ackermann}, M., {Ajello}, M., {Albert}, A., {et~al.} 2014, Science, 345, 554

\bibitem[{{Ackermann} {et~al.}(2015){Ackermann}, {Arcavi}, {Baldini}, {Ballet}, {Barbiellini}, {Bastieri}, {Bellazzini}, {Bissaldi}, {Blandford}, {Bonino}, {Bottacini}, {Brandt}, {Bregeon}, {Bruel}, {Buehler}, {Buson}, {Caliandro}, {Cameron}, {Caragiulo}, {Caraveo}, {Cavazzuti}, {Cecchi}, {Charles}, {Chekhtman}, {Chiang}, {Chiaro}, {Ciprini}, {Claus}, {Cohen-Tanugi}, {Cutini}, {D'Ammando}, {de Angelis}, {de Palma}, {Desiante}, {Di Venere}, {Drell}, {Favuzzi}, {Fegan}, {Franckowiak}, {Funk}, {Fusco}, {Gal-Yam}, {Gargano}, {Gasparrini}, {Giglietto}, {Giordano}, {Giroletti}, {Glanzman}, {Godfrey}, {Grenier}, {Grove}, {Guiriec}, {Harding}, {Hayashi}, {Hewitt}, {Hill}, {Horan}, {Jogler}, {J{\'o}hannesson}, {Kocevski}, {Kuss}, {Larsson}, {Lashner}, {Latronico}, {Li}, {Li}, {Longo}, {Loparco}, {Lovellette}, {Lubrano}, {Malyshev}, {Mayer}, {Mazziotta}, {McEnery}, {Michelson}, {Mizuno}, {Monzani}, {Morselli}, {Murase}, {Nugent}, {Nuss}, {Ofek}, {Ohsugi}, {Orienti}, {Orlando}, {Ormes}, {Paneque}, {Pesce-Rollins},
  {Piron}, {Pivato}, {Rain{\`o}}, {Rando}, {Razzano}, {Reimer}, {Reimer}, {Schulz}, {Sgr{\`o}}, {Siskind}, {Spada}, {Spandre}, {Spinelli}, {Suson}, {Takahashi}, {Thayer}, {Tibaldo}, {Torres}, {Troja}, {Vianello}, {Werner}, {Wood}, \& {Wood}}]{2015ApJ...807..169A}
{Ackermann}, M., {Arcavi}, I., {Baldini}, L., {et~al.} 2015, \apj, 807, 169

\bibitem[{{Astropy Collaboration} {et~al.}(2018){Astropy Collaboration}, {Price-Whelan}, {Sip{\H{o}}cz}, {G{\"u}nther}, {Lim}, {Crawford}, {Conseil}, {Shupe}, {Craig}, {Dencheva}, {Ginsburg}, {Vand erPlas}, {Bradley}, {P{\'e}rez-Su{\'a}rez}, {de Val-Borro}, {Aldcroft}, {Cruz}, {Robitaille}, {Tollerud}, {Ardelean}, {Babej}, {Bach}, {Bachetti}, {Bakanov}, {Bamford}, {Barentsen}, {Barmby}, {Baumbach}, {Berry}, {Biscani}, {Boquien}, {Bostroem}, {Bouma}, {Brammer}, {Bray}, {Breytenbach}, {Buddelmeijer}, {Burke}, {Calderone}, {Cano Rodr{\'\i}guez}, {Cara}, {Cardoso}, {Cheedella}, {Copin}, {Corrales}, {Crichton}, {D'Avella}, {Deil}, {Depagne}, {Dietrich}, {Donath}, {Droettboom}, {Earl}, {Erben}, {Fabbro}, {Ferreira}, {Finethy}, {Fox}, {Garrison}, {Gibbons}, {Goldstein}, {Gommers}, {Greco}, {Greenfield}, {Groener}, {Grollier}, {Hagen}, {Hirst}, {Homeier}, {Horton}, {Hosseinzadeh}, {Hu}, {Hunkeler}, {Ivezi{\'c}}, {Jain}, {Jenness}, {Kanarek}, {Kendrew}, {Kern}, {Kerzendorf}, {Khvalko}, {King}, {Kirkby}, {Kulkarni},
  {Kumar}, {Lee}, {Lenz}, {Littlefair}, {Ma}, {Macleod}, {Mastropietro}, {McCully}, {Montagnac}, {Morris}, {Mueller}, {Mumford}, {Muna}, {Murphy}, {Nelson}, {Nguyen}, {Ninan}, {N{\"o}the}, {Ogaz}, {Oh}, {Parejko}, {Parley}, {Pascual}, {Patil}, {Patil}, {Plunkett}, {Prochaska}, {Rastogi}, {Reddy Janga}, {Sabater}, {Sakurikar}, {Seifert}, {Sherbert}, {Sherwood-Taylor}, {Shih}, {Sick}, {Silbiger}, {Singanamalla}, {Singer}, {Sladen}, {Sooley}, {Sornarajah}, {Streicher}, {Teuben}, {Thomas}, {Tremblay}, {Turner}, {Terr{\'o}n}, {van Kerkwijk}, {de la Vega}, {Watkins}, {Weaver}, {Whitmore}, {Woillez}, {Zabalza}, \& {Astropy Contributors}}]{astropy:2018}
{Astropy Collaboration}, {Price-Whelan}, A.~M., {Sip{\H{o}}cz}, B.~M., {et~al.} 2018, \aj, 156, 123

\bibitem[{{Baade} \& {Zwicky}(1934)}]{1934PNAS...20..259B}
{Baade}, W. \& {Zwicky}, F. 1934, Proceedings of the National Academy of Science, 20, 259

\bibitem[{{Bell} {et~al.}(2013){Bell}, {Schure}, {Reville}, \& {Giacinti}}]{2013MNRAS.431..415B}
{Bell}, A.~R., {Schure}, K.~M., {Reville}, B., \& {Giacinti}, G. 2013, \mnras, 431, 415

\bibitem[{{Bellm} {et~al.}(2019){Bellm}, {Kulkarni}, {Graham}, {Dekany}, {Smith}, {Riddle}, {Masci}, {Helou}, {Prince}, {Adams}, {Barbarino}, {Barlow}, {Bauer}, {Beck}, {Belicki}, {Biswas}, {Blagorodnova}, {Bodewits}, {Bolin}, {Brinnel}, {Brooke}, {Bue}, {Bulla}, {Burruss}, {Cenko}, {Chang}, {Connolly}, {Coughlin}, {Cromer}, {Cunningham}, {De}, {Delacroix}, {Desai}, {Duev}, {Eadie}, {Farnham}, {Feeney}, {Feindt}, {Flynn}, {Franckowiak}, {Frederick}, {Fremling}, {Gal-Yam}, {Gezari}, {Giomi}, {Goldstein}, {Golkhou}, {Goobar}, {Groom}, {Hacopians}, {Hale}, {Henning}, {Ho}, {Hover}, {Howell}, {Hung}, {Huppenkothen}, {Imel}, {Ip}, {Ivezi{\'c}}, {Jackson}, {Jones}, {Juric}, {Kasliwal}, {Kaspi}, {Kaye}, {Kelley}, {Kowalski}, {Kramer}, {Kupfer}, {Landry}, {Laher}, {Lee}, {Lin}, {Lin}, {Lunnan}, {Giomi}, {Mahabal}, {Mao}, {Miller}, {Monkewitz}, {Murphy}, {Ngeow}, {Nordin}, {Nugent}, {Ofek}, {Patterson}, {Penprase}, {Porter}, {Rauch}, {Rebbapragada}, {Reiley}, {Rigault}, {Rodriguez}, {van Roestel}, {Rusholme}, {van
  Santen}, {Schulze}, {Shupe}, {Singer}, {Soumagnac}, {Stein}, {Surace}, {Sollerman}, {Szkody}, {Taddia}, {Terek}, {Van Sistine}, {van Velzen}, {Vestrand}, {Walters}, {Ward}, {Ye}, {Yu}, {Yan}, \& {Zolkower}}]{2019PASP..131a8002B}
{Bellm}, E.~C., {Kulkarni}, S.~R., {Graham}, M.~J., {et~al.} 2019, \pasp, 131, 018002

\bibitem[{{Bhatt} {et~al.}(2020){Bhatt}, {Sushch}, {Pohl}, {Fedynitch}, {Das}, {Brose}, {Plotko}, \& {Meyer}}]{2020APh...12302490B}
{Bhatt}, M., {Sushch}, I., {Pohl}, M., {et~al.} 2020, Astroparticle Physics, 123, 102490

\bibitem[{{Bietenholz} {et~al.}(2021){Bietenholz}, {Bartel}, {Argo}, {Dua}, {Ryder}, \& {Soderberg}}]{2021ApJ...908...75B}
{Bietenholz}, M.~F., {Bartel}, N., {Argo}, M., {et~al.} 2021, \apj, 908, 75

\bibitem[{{Blandford} \& {Eichler}(1987)}]{1987PhR...154....1B}
{Blandford}, R. \& {Eichler}, D. 1987, Physics Reports, 154, 1

\bibitem[{{Brose} {et~al.}(2021){Brose}, {Pohl}, \& {Sushch}}]{2021A&A...654A.139B}
{Brose}, R., {Pohl}, M., \& {Sushch}, I. 2021, \aap, 654, A139

\bibitem[{{Brose} {et~al.}(2022){Brose}, {Sushch}, \& {Mackey}}]{2022MNRAS.516..492B}
{Brose}, R., {Sushch}, I., \& {Mackey}, J. 2022, \mnras, 516, 492

\bibitem[{{Brose} {et~al.}(2025){Brose}, {Sushch}, \& {Mackey}}]{Paper2}
{Brose}, R., {Sushch}, I., \& {Mackey}, J. 2025, \mnras, 516, 492

\bibitem[{{Brose} {et~al.}(2016){Brose}, {Telezhinsky}, \& {Pohl}}]{2016A&A...593A..20B}
{Brose}, R., {Telezhinsky}, I., \& {Pohl}, M. 2016, \aap, 593, A20

\bibitem[{{Bruch} {et~al.}(2021){Bruch}, {Gal-Yam}, {Schulze}, {Yaron}, {Yang}, {Soumagnac}, {Rigault}, {Strotjohann}, {Ofek}, {Sollerman}, {Masci}, {Barbarino}, {Ho}, {Fremling}, {Perley}, {Nordin}, {Cenko}, {Adams}, {Adreoni}, {Bellm}, {Blagorodnova}, {Bulla}, {Burdge}, {De}, {Dhawan}, {Drake}, {Duev}, {Dugas}, {Graham}, {Graham}, {Irani}, {Jencson}, {Karamehmetoglu}, {Kasliwal}, {Kim}, {Kulkarni}, {Kupfer}, {Liang}, {Mahabal}, {Miller}, {Prince}, {Riddle}, {Sharma}, {Smith}, {Taddia}, {Taggart}, {Walters}, \& {Yan}}]{2021ApJ...912...46B}
{Bruch}, R.~J., {Gal-Yam}, A., {Schulze}, S., {et~al.} 2021, \apj, 912, 46

\bibitem[{{Cherenkov Telescope Array Observatory and Cherenkov Telescope Array Consortium}(2021)}]{cherenkov_telescope_array_observatory_2021_5499840}
{Cherenkov Telescope Array Observatory and Cherenkov Telescope Array Consortium}. 2021, CTAO Instrument Response Functions - prod5 version v0.1

\bibitem[{{Chevalier} \& {Fransson}(2006)}]{2006ApJ...651..381C}
{Chevalier}, R.~A. \& {Fransson}, C. 2006, \apj, 651, 381

\bibitem[{{Chevalier} {et~al.}(2006){Chevalier}, {Fransson}, \& {Nymark}}]{2006ApJ...641.1029C}
{Chevalier}, R.~A., {Fransson}, C., \& {Nymark}, T.~K. 2006, \apj, 641, 1029

\bibitem[{{Cold} \& {Hjorth}(2023)}]{2023A&A...670A..48C}
{Cold}, C. \& {Hjorth}, J. 2023, \aap, 670, A48

\bibitem[{{Cristofari} {et~al.}(2020{\natexlab{a}}){Cristofari}, {Blasi}, \& {Amato}}]{2020APh...12302492C}
{Cristofari}, P., {Blasi}, P., \& {Amato}, E. 2020{\natexlab{a}}, Astroparticle Physics, 123, 102492

\bibitem[{{Cristofari} {et~al.}(2022){Cristofari}, {Marcowith}, {Renaud}, {Dwarkadas}, {Tatischeff}, {Giacinti}, {Peretti}, \& {Sol}}]{2022MNRAS.511.3321C}
{Cristofari}, P., {Marcowith}, A., {Renaud}, M., {et~al.} 2022, \mnras, 511, 3321

\bibitem[{{Cristofari} {et~al.}(2020{\natexlab{b}}){Cristofari}, {Renaud}, {Marcowith}, {Dwarkadas}, \& {Tatischeff}}]{2020MNRAS.494.2760C}
{Cristofari}, P., {Renaud}, M., {Marcowith}, A., {Dwarkadas}, V.~V., \& {Tatischeff}, V. 2020{\natexlab{b}}, \mnras, 494, 2760

\bibitem[{Dastidar {et~al.}(2018)Dastidar, Misra, Hosseinzadeh, Pastorello, Pumo, Valenti, McCully, Tomasella, Arcavi, Elias-Rosa, Singh, Gangopadhyay, Howell, Morales-Garoffolo, Zampieri, Kumar, Turatto, Benetti, Tartaglia, Ochner, Sahu, Anupama, \& Pandey}]{10.1093/mnras/sty1634}
Dastidar, R., Misra, K., Hosseinzadeh, G., {et~al.} 2018, Monthly Notices of the Royal Astronomical Society, 479, 2421

\bibitem[{{De Angelis} {et~al.}(2017){De Angelis}, {Tatischeff}, {Tavani}, {Oberlack}, {Grenier}, {Hanlon}, {Walter}, {Argan}, {von Ballmoos}, {Bulgarelli}, {Donnarumma}, {Hernanz}, {Kuvvetli}, {Pearce}, {Zdziarski}, {Aboudan}, {Ajello}, {Ambrosi}, {Bernard}, {Bernardini}, {Bonvicini}, {Brogna}, {Branchesi}, {Budtz-Jorgensen}, {Bykov}, {Campana}, {Cardillo}, {Coppi}, {De Martino}, {Diehl}, {Doro}, {Fioretti}, {Funk}, {Ghisellini}, {Grove}, {Hamadache}, {Hartmann}, {Hayashida}, {Isern}, {Kanbach}, {Kiener}, {Kn{\"o}dlseder}, {Labanti}, {Laurent}, {Limousin}, {Longo}, {Mannheim}, {Marisaldi}, {Martinez}, {Mazziotta}, {McEnery}, {Mereghetti}, {Minervini}, {Moiseev}, {Morselli}, {Nakazawa}, {Orleanski}, {Paredes}, {Patricelli}, {Peyr{\'e}}, {Piano}, {Pohl}, {Ramarijaona}, {Rando}, {Reichardt}, {Roncadelli}, {Silva}, {Tavecchio}, {Thompson}, {Turolla}, {Ulyanov}, {Vacchi}, {Wu}, \& {Zoglauer}}]{2017ExA....44...25D}
{De Angelis}, A., {Tatischeff}, V., {Tavani}, M., {et~al.} 2017, Experimental Astronomy, 44, 25

\bibitem[{{de O{\~n}a Wilhelmi} {et~al.}(2020){de O{\~n}a Wilhelmi}, {Sushch}, {Brose}, {Mestre}, {Su}, \& {Zanin}}]{2020MNRAS.497.3581D}
{de O{\~n}a Wilhelmi}, E., {Sushch}, I., {Brose}, R., {et~al.} 2020, \mnras, 497, 3581

\bibitem[{{Diesing}(2023)}]{2023ApJ...958....3D}
{Diesing}, R. 2023, \apj, 958, 3

\bibitem[{{Dubus}(2006)}]{2006A&A...451....9D}
{Dubus}, G. 2006, \aap, 451, 9

\bibitem[{{Dwarkadas}(2014)}]{2014MNRAS.440.1917D}
{Dwarkadas}, V.~V. 2014, \mnras, 440, 1917

\bibitem[{{Dwarkadas}(2025)}]{2025arXiv250508946D}
{Dwarkadas}, V.~V. 2025, arXiv e-prints, arXiv:2505.08946

\bibitem[{{Finke} {et~al.}(2010){Finke}, {Razzaque}, \& {Dermer}}]{2010ApJ...712..238F}
{Finke}, J.~D., {Razzaque}, S., \& {Dermer}, C.~D. 2010, \apj, 712, 238

\bibitem[{{Fox} {et~al.}(2013){Fox}, {Filippenko}, {Skrutskie}, {Silverman}, {Ganeshalingam}, {Cenko}, \& {Clubb}}]{2013AJ....146....2F}
{Fox}, O.~D., {Filippenko}, A.~V., {Skrutskie}, M.~F., {et~al.} 2013, \aj, 146, 2

\bibitem[{{Gal-Yam}(2021)}]{2021AAS...23742305G}
{Gal-Yam}, A. 2021, in American Astronomical Society Meeting Abstracts, Vol. 237, American Astronomical Society Meeting Abstracts \#237, 423.05

\bibitem[{{Gould} \& {Schr{\'e}der}(1966)}]{1966PhRvL..16..252G}
{Gould}, R.~J. \& {Schr{\'e}der}, G. 1966, \prl, 16, 252

\bibitem[{{Gould} \& {Schr{\'e}der}(1967)}]{1967PhRv..155.1404G}
{Gould}, R.~J. \& {Schr{\'e}der}, G.~P. 1967, Physical Review, 155, 1404

\bibitem[{{Grassitelli} {et~al.}(2021){Grassitelli}, {Langer}, {Mackey}, {Gr{\"a}fener}, {Grin}, {Sander}, \& {Vink}}]{2021A&A...647A..99G}
{Grassitelli}, L., {Langer}, N., {Mackey}, J., {et~al.} 2021, \aap, 647, A99

\bibitem[{{H.E.S.S. Collaboration} {et~al.}(2019){H.E.S.S. Collaboration}, {Abdalla}, {Aharonian}, {Ait Benkhali}, {Ang{\"u}ner}, {Arakawa}, {Arcaro}, {Armand}, {Ashkar}, {Backes}, {Barbosa Martins}, {Barnard}, {Becherini}, {Berge}, {Bernl{\"o}hr}, {Blackwell}, {B{\"o}ttcher}, {Boisson}, {Bolmont}, {Bonnefoy}, {Bregeon}, {Breuhaus}, {Brun}, {Brun}, {Bryan}, {B{\"u}chele}, {Bulik}, {Bylund}, {Capasso}, {Caroff}, {Carosi}, {Casanova}, {Cerruti}, {Chakraborty}, {Chand}, {Chandra}, {Chaves}, {Chen}, {Colafrancesco}, {Curylo}, {Davids}, {Deil}, {Devin}, {de Wilt}, {Dirson}, {Djannati-Ata{\"\i}}, {Dmytriiev}, {Donath}, {Doroshenko}, {Drury}, {Dyks}, {Egberts}, {Emery}, {Ernenwein}, {Eschbach}, {Feijen}, {Fegan}, {Fiasson}, {Fontaine}, {Funk}, {F{\"u}{\ss}ling}, {Gabici}, {Gallant}, {Gat{\'e}}, {Giavitto}, {Glawion}, {Glicenstein}, {Gottschall}, {Grondin}, {Hahn}, {Haupt}, {Heinzelmann}, {Henri}, {Hermann}, {Hinton}, {Hofmann}, {Hoischen}, {Holch}, {Holler}, {Horns}, {Huber}, {Iwasaki}, {Jamrozy}, {Jankowsky},
  {Jankowsky}, {Jung-Richardt}, {Kastendieck}, {Katarzy{\'n}ski}, {Katsuragawa}, {Katz}, {Khangulyan}, {Kh{\'e}lifi}, {King}, {Klepser}, {Klu{\'z}niak}, {Komin}, {Kosack}, {Kostunin}, {Kraus}, {Lamanna}, {Lau}, {Lemi{\`e}re}, {Lemoine-Goumard}, {Lenain}, {Leser}, {Levy}, {Lohse}, {L{\'o}pez-Coto}, {Lypova}, {Mackey}, {Majumdar}, {Malyshev}, {Marandon}, {Marcowith}, {Mares}, {Mariaud}, {Mart{\'\i}-Devesa}, {Marx}, {Maurin}, {Meintjes}, {Mitchell}, {Moderski}, {Mohamed}, {Mohrmann}, {Muller}, {Moore}, {Moulin}, {Murach}, {Nakashima}, {de Naurois}, {Ndiyavala}, {Niederwanger}, {Niemiec}, {Oakes}, {O'Brien}, {Odaka}, {Ohm}, {de Ona Wilhelmi}, {Ostrowski}, {Oya}, {Panter}, {Parsons}, {Perennes}, {Petrucci}, {Peyaud}, {Piel}, {Pita}, {Poireau}, {Priyana Noel}, {Prokhorov}, {Prokoph}, {P{\"u}hlhofer}, {Punch}, {Quirrenbach}, {Raab}, {Rauth}, {Reimer}, {Reimer}, {Remy}, {Renaud}, {Rieger}, {Rinchiuso}, {Romoli}, {Rowell}, {Rudak}, {Ruiz-Velasco}, {Sahakian}, {Saito}, {Sanchez}, {Santangelo}, {Sasaki}, {Schlickeiser},
  {Sch{\"u}ssler}, {Schulz}, {Schutte}, {Schwanke}, {Schwemmer}, {Seglar-Arroyo}, {Senniappan}, {Seyffert}, {Shafi}, {Shiningayamwe}, {Simoni}, {Sinha}, {Sol}, {Specovius}, {Spir-Jacob}, {Stawarz}, {Steenkamp}, {Stegmann}, {Steppa}, {Takahashi}, {Tavernier}, {Taylor}, {Terrier}, {Tiziani}, {Tluczykont}, {Trichard}, {Tsirou}, {Tsuji}, {Tuffs}, {Uchiyama}, {van der Walt}, {van Eldik}, {van Rensburg}, {van Soelen}, {Vasileiadis}, {Veh}, {Venter}, {Vincent}, {Vink}, {Voisin}, {V{\"o}lk}, {Vuillaume}, {Wadiasingh}, {Wagner}, {White}, {Wierzcholska}, {Yang}, {Yoneda}, {Zacharias}, {Zanin}, {Zdziarski}, {Zech}, {Ziegler}, {Zorn}, {{\.Z}ywucka}, \& {Maxted}}]{2019A&A...626A..57H}
{H.E.S.S. Collaboration}, {Abdalla}, H., {Aharonian}, F., {et~al.} 2019, \aap, 626, A57

\bibitem[{{H.E.S.S. Collaboration} {et~al.}(2022){H.E.S.S. Collaboration}, {Aharonian}, {Ait Benkhali}, {Ang{\"u}ner}, {Ashkar}, {Backes}, {Baghmanyan}, {Barbosa Martins}, {Batzofin}, {Becherini}, {Berge}, {Bernl{\"o}hr}, {Bi}, {B{\"o}ttcher}, {Boisson}, {Bolmont}, {de Bony de Lavergne}, {Breuhaus}, {Brose}, {Brun}, {Caroff}, {Casanova}, {Cerruti}, {Chand}, {Chen}, {Cotter}, {Damascene Mbarubucyeye}, {Djannati-Ata{\"\i}}, {Dmytriiev}, {Doroshenko}, {Duffy}, {Egberts}, {Ernenwein}, {Fegan}, {Feijen}, {Fiasson}, {Fichet de Clairfontaine}, {Fontaine}, {F{\"u}{\ss}ling}, {Funk}, {Gabici}, {Gallant}, {Ghafourizadeh}, {Giavitto}, {Giunti}, {Glawion}, {Glicenstein}, {Grondin}, {Hermann}, {Hinton}, {H{\"o}rbe}, {Hofmann}, {Hoischen}, {Holch}, {Holler}, {Horns}, {Huang}, {Jamrozy}, {Jankowsky}, {Jung-Richardt}, {Kasai}, {Katarzy{\'n}ski}, {Katz}, {Khangulyan}, {Kh{\'e}lifi}, {Klepser}, {Klu{\'z}niak}, {Komin}, {Konno}, {Kosack}, {Kostunin}, {Le Stum}, {Lemi{\`e}re}, {Lemoine-Goumard}, {Lenain}, {Leuschner}, {Lohse},
  {Luashvili}, {Lypova}, {Mackey}, {Malyshev}, {Malyshev}, {Marandon}, {Marchegiani}, {Marcowith}, {Mart{\'\i}-Devesa}, {Marx}, {Maurin}, {Meyer}, {Mitchell}, {Moderski}, {Mohrmann}, {Montanari}, {Moulin}, {Muller}, {Murach}, {Nakashima}, {de Naurois}, {Nayerhoda}, {Niemiec}, {Priyana Noel}, {O{\textquoteright}Brien}, {Ohm}, {Olivera-Nieto}, {de Ona Wilhelmi}, {Ostrowski}, {Panny}, {Panter}, {Parsons}, {Peron}, {Pita}, {Poireau}, {Prokhorov}, {Prokoph}, {P{\"u}hlhofer}, {Punch}, {Quirrenbach}, {Reichherzer}, {Reimer}, {Reimer}, {Renaud}, {Reville}, {Rieger}, {Rowell}, {Rudak}, {Rueda Ricarte}, {Ruiz-Velasco}, {Sahakian}, {Sailer}, {Salzmann}, {Sanchez}, {Santangelo}, {Sasaki}, {Sch{\"a}fer}, {Sch{\"u}ssler}, {Schutte}, {Schwanke}, {Senniappan}, {Shapopi}, {Simoni}, {Sinha}, {Sol}, {Specovius}, {Spencer}, {Stawarz}, {Steinmassl}, {Steppa}, {Takahashi}, {Tanaka}, {Taylor}, {Terrier}, {Thorpe-Morgan}, {Tsirou}, {Tsuji}, {Tuffs}, {Uchiyama}, {Unbehaun}, {van Eldik}, {van Soelen}, {Veh}, {Venter}, {Vink},
  {Wagner}, {Werner}, {White}, {Wierzcholska}, {Wong}, {Yusafzai}, {Zacharias}, {Zargaryan}, {Zdziarski}, {Zech}, {Zhu}, {Zouari}, \& {{\.Z}ywucka}}]{HESS2022_RSOph}
{H.E.S.S. Collaboration}, {Aharonian}, F., {Ait Benkhali}, F., {et~al.} 2022, Science, 376, 77

\bibitem[{{Hnatyk} \& {Petruk}(1999)}]{1999A&A...344..295H}
{Hnatyk}, B. \& {Petruk}, O. 1999, \aap, 344, 295

\bibitem[{{Hodapp} {et~al.}(2004){Hodapp}, {Kaiser}, {Aussel}, {Burgett}, {Chambers}, {Chun}, {Dombeck}, {Douglas}, {Hafner}, {Heasley}, {Hoblitt}, {Hude}, {Isani}, {Jedicke}, {Jewitt}, {Laux}, {Luppino}, {Lupton}, {Maberry}, {Magnier}, {Mannery}, {Monet}, {Morgan}, {Onaka}, {Price}, {Ryan}, {Siegmund}, {Szapudi}, {Tonry}, {Wainscoat}, \& {Waterson}}]{2004AN....325..636H}
{Hodapp}, K.~W., {Kaiser}, N., {Aussel}, H., {et~al.} 2004, Astronomische Nachrichten, 325, 636

\bibitem[{{Hodgkin} {et~al.}(2013){Hodgkin}, {Wyrzykowski}, {Blagorodnova}, \& {Koposov}}]{2013RSPTA.37120239H}
{Hodgkin}, S.~T., {Wyrzykowski}, L., {Blagorodnova}, N., \& {Koposov}, S. 2013, Philosophical Transactions of the Royal Society of London Series A, 371, 20120239

\bibitem[{{Huang} {et~al.}(2007){Huang}, {Park}, {Pohl}, \& {Daniels}}]{2007APh....27..429H}
{Huang}, C.~Y., {Park}, S.~E., {Pohl}, M., \& {Daniels}, C.~D. 2007, Astroparticle Physics, 27, 429

\bibitem[{{Inoue} {et~al.}(2021){Inoue}, {Marcowith}, {Giacinti}, {Jan van Marle}, \& {Nishino}}]{2021ApJ...922....7I}
{Inoue}, T., {Marcowith}, A., {Giacinti}, G., {Jan van Marle}, A., \& {Nishino}, S. 2021, \apj, 922, 7

\bibitem[{{Jacobson-Gal{\'a}n} {et~al.}(2025){Jacobson-Gal{\'a}n}, {Dessart}, {Davis}, {Bostroem}, {Kilpatrick}, {Margutti}, {Filippenko}, {Foley}, {Chornock}, {Terreran}, {Hiramatsu}, {Newsome}, {Padilla Gonzalez}, {Pellegrino}, {Howell}, {Anderson}, {Angus}, {Auchettl}, {Brink}, {Cartier}, {Coulter}, {de Boer}, {Drout}, {Earl}, {Ertini}, {Farah}, {Farias}, {Gall}, {Gao}, {Gerlach}, {Guo}, {Haynie}, {Hosseinzadeh}, {Ibik}, {Jha}, {Jones}, {Langeroodi}, {LeBaron}, {Magnier}, {Piro}, {Raimundo}, {Rest}, {Rest}, {Rich}, {Rojas-Bravo}, {Sears}, {Taggart}, {Villar}, {Wainscoat}, {Wang}, {Wasserman}, {Yan}, {Yang}, {Zhang}, \& {Zheng}}]{2025arXiv250504698J}
{Jacobson-Gal{\'a}n}, W.~V., {Dessart}, L., {Davis}, K.~W., {et~al.} 2025, arXiv e-prints, arXiv:2505.04698

\bibitem[{{Jacobson-Gal{\'a}n} {et~al.}(2022){Jacobson-Gal{\'a}n}, {Dessart}, {Jones}, {Margutti}, {Coppejans}, {Dimitriadis}, {Foley}, {Kilpatrick}, {Matthews}, {Rest}, {Terreran}, {Aleo}, {Auchettl}, {Blanchard}, {Coulter}, {Davis}, {de Boer}, {DeMarchi}, {Drout}, {Earl}, {Gagliano}, {Gall}, {Hjorth}, {Huber}, {Ibik}, {Milisavljevic}, {Pan}, {Rest}, {Ridden-Harper}, {Rojas-Bravo}, {Siebert}, {Smith}, {Taggart}, {Tinyanont}, {Wang}, \& {Zenati}}]{2022ApJ...924...15J}
{Jacobson-Gal{\'a}n}, W.~V., {Dessart}, L., {Jones}, D.~O., {et~al.} 2022, \apj, 924, 15

\bibitem[{{Jauch} \& {Rohrlich}(1976)}]{1976tper.book.....J}
{Jauch}, J.~M. \& {Rohrlich}, F. 1976, {The theory of photons and electrons. The relativistic quantum field theory of charged particles with spin one-half}

\bibitem[{{Jonas} \& {MeerKAT Team}(2016)}]{jonas16}
{Jonas}, J. \& {MeerKAT Team}. 2016, in MeerKAT Science: On the Pathway to the SKA, 1

\bibitem[{{Kozyreva} {et~al.}(2025){Kozyreva}, {Caputo}, {Baklanov}, {Mironov}, \& {Janka}}]{2025A&A...694A.319K}
{Kozyreva}, A., {Caputo}, A., {Baklanov}, P., {Mironov}, A., \& {Janka}, H.-T. 2025, \aap, 694, A319

\bibitem[{{Langer}(2012)}]{Lan12}
{Langer}, N. 2012, \araa, 50, 107

\bibitem[{{Lundqvist} \& {Fransson}(1996)}]{1996ApJ...464..924L}
{Lundqvist}, P. \& {Fransson}, C. 1996, \apj, 464, 924

\bibitem[{{Mackey} {et~al.}(2015){Mackey}, {Castro}, {Fossati}, \& {Langer}}]{2015A&A...582A..24M}
{Mackey}, J., {Castro}, N., {Fossati}, L., \& {Langer}, N. 2015, \aap, 582, A24

\bibitem[{{Mackey} {et~al.}(2021){Mackey}, {Green}, {Moutzouri}, {Haworth}, {Kavanagh}, {Zargaryan}, \& {Celeste}}]{2021MNRAS.504..983M}
{Mackey}, J., {Green}, S., {Moutzouri}, M., {et~al.} 2021, \mnras, 504, 983

\bibitem[{{Mackey} {et~al.}(2014){Mackey}, {Mohamed}, {Gvaramadze}, {Kotak}, {Langer}, {Meyer}, {Moriya}, \& {Neilson}}]{2014Natur.512..282M}
{Mackey}, J., {Mohamed}, S., {Gvaramadze}, V.~V., {et~al.} 2014, \nat, 512, 282

\bibitem[{{Mackey} {et~al.}(2019){Mackey}, {Walch}, {Seifried}, {Glover}, {W{\"u}nsch}, \& {Aharonian}}]{2019MNRAS.486.1094M}
{Mackey}, J., {Walch}, S., {Seifried}, D., {et~al.} 2019, \mnras, 486, 1094

\bibitem[{{Marcowith} {et~al.}(2018){Marcowith}, {Dwarkadas}, {Renaud}, {Tatischeff}, \& {Giacinti}}]{2018MNRAS.479.4470M}
{Marcowith}, A., {Dwarkadas}, V.~V., {Renaud}, M., {Tatischeff}, V., \& {Giacinti}, G. 2018, \mnras, 479, 4470

\bibitem[{{Mart{\'\i}-Devesa} {et~al.}(2024){Mart{\'\i}-Devesa}, {Cheung}, {Di Lalla}, {Renaud}, {Principe}, {Omodei}, \& {Acero}}]{2024A&A...686A.254M}
{Mart{\'\i}-Devesa}, G., {Cheung}, C.~C., {Di Lalla}, N., {et~al.} 2024, \aap, 686, A254

\bibitem[{{Marti-Devesa} \& {Fermi-LAT Collaboration}(2024)}]{2024ATel16601....1M}
{Marti-Devesa}, G. \& {Fermi-LAT Collaboration}. 2024, The Astronomer's Telegram, 16601, 1

\bibitem[{{Matsuoka} {et~al.}(2025){Matsuoka}, {Maeda}, {Kimura}, \& {Tanaka}}]{2025arXiv250506609M}
{Matsuoka}, T., {Maeda}, K., {Kimura}, S.~S., \& {Tanaka}, M. 2025, arXiv e-prints, arXiv:2505.06609

\bibitem[{{Mezger} \& {Henderson}(1967)}]{1967ApJ...147..471M}
{Mezger}, P.~G. \& {Henderson}, A.~P. 1967, \apj, 147, 471

\bibitem[{{Mignone} {et~al.}(2007){Mignone}, {Bodo}, {Massaglia}, {Matsakos}, {Tesileanu}, {Zanni}, \& {Ferrari}}]{2007ApJS..170..228M}
{Mignone}, A., {Bodo}, G., {Massaglia}, S., {et~al.} 2007, \apjs, 170, 228

\bibitem[{{Milisavljevic} {et~al.}(2015){Milisavljevic}, {Margutti}, {Kamble}, {Patnaude}, {Raymond}, {Eldridge}, {Fong}, {Bietenholz}, {Challis}, {Chornock}, {Drout}, {Fransson}, {Fesen}, {Grindlay}, {Kirshner}, {Lunnan}, {Mackey}, {Miller}, {Parrent}, {Sanders}, {Soderberg}, \& {Zauderer}}]{2015ApJ...815..120M}
{Milisavljevic}, D., {Margutti}, R., {Kamble}, A., {et~al.} 2015, \apj, 815, 120

\bibitem[{{Moran} {et~al.}(2023){Moran}, {Fraser}, {Kotak}, {Pastorello}, {Benetti}, {Brennan}, {Guti{\'e}rrez}, {Kankare}, {Kuncarayakti}, {Mattila}, {Reynolds}, {Anderson}, {Brown}, {Campana}, {Chambers}, {Chen}, {Della Valle}, {Dennefeld}, {Elias-Rosa}, {Galbany}, {Galindo-Guil}, {Gromadzki}, {Hiramatsu}, {Inserra}, {Leloudas}, {M{\"u}ller-Bravo}, {Nicholl}, {Reguitti}, {Shahbandeh}, {Smartt}, {Tartaglia}, \& {Young}}]{2023A&A...669A..51M}
{Moran}, S., {Fraser}, M., {Kotak}, R., {et~al.} 2023, \aap, 669, A51

\bibitem[{{Moriya} {et~al.}(2013){Moriya}, {Blinnikov}, {Tominaga}, {Yoshida}, {Tanaka}, {Maeda}, \& {Nomoto}}]{2013MNRAS.428.1020M}
{Moriya}, T.~J., {Blinnikov}, S.~I., {Tominaga}, N., {et~al.} 2013, \mnras, 428, 1020

\bibitem[{{Morrison} \& {McCammon}(1983)}]{1983ApJ...270..119M}
{Morrison}, R. \& {McCammon}, D. 1983, \apj, 270, 119

\bibitem[{{Murase} {et~al.}(2011){Murase}, {Thompson}, {Lacki}, \& {Beacom}}]{MurThoLac11}
{Murase}, K., {Thompson}, T.~A., {Lacki}, B.~C., \& {Beacom}, J.~F. 2011, \prd, 84, 043003

\bibitem[{{Murase} {et~al.}(2014){Murase}, {Thompson}, \& {Ofek}}]{MurThoOfe14}
{Murase}, K., {Thompson}, T.~A., \& {Ofek}, E.~O. 2014, \mnras, 440, 2528

\bibitem[{{Prokhorov} {et~al.}(2021){Prokhorov}, {Moraghan}, \& {Vink}}]{2021MNRAS.tmp.1282P}
{Prokhorov}, D.~A., {Moraghan}, A., \& {Vink}, J. 2021, \mnras [\eprint[arXiv]{2105.04007}]

\bibitem[{{Shrestha} {et~al.}(2024){Shrestha}, {Bostroem}, {Sand}, {Hosseinzadeh}, {Andrews}, {Dong}, {Hoang}, {Janzen}, {Pearson}, {Jencson}, {Lundquist}, {Mehta}, {Ravi}, {Meza Retamal}, {Valenti}, {Brown}, {Jha}, {Macrie}, {Hsu}, {Farah}, {Howell}, {McCully}, {Newsome}, {Padilla Gonzalez}, {Pellegrino}, {Terreran}, {Kwok}, {Smith}, {Schwab}, {Martas}, {Munoz}, {Medina}, {Li}, {Diaz}, {Hiramatsu}, {Tucker}, {Wheeler}, {Wang}, {Zhai}, {Zhang}, {Gangopadhyay}, {Yang}, \& {Guti{\'e}rrez}}]{2024ApJ...972L..15S}
{Shrestha}, M., {Bostroem}, K.~A., {Sand}, D.~J., {et~al.} 2024, \apjl, 972, L15

\bibitem[{{Shvartzvald} {et~al.}(2024){Shvartzvald}, {Waxman}, {Gal-Yam}, {Ofek}, {Ben-Ami}, {Berge}, {Kowalski}, {B{\"u}hler}, {Worm}, {Rhoads}, {Arcavi}, {Maoz}, {Polishook}, {Stone}, {Trakhtenbrot}, {Ackermann}, {Aharonson}, {Birnholtz}, {Chelouche}, {Guetta}, {Hallakoun}, {Horesh}, {Kushnir}, {Mazeh}, {Nordin}, {Ofir}, {Ohm}, {Parsons}, {Pe'er}, {Perets}, {Perdelwitz}, {Poznanski}, {Sadeh}, {Sagiv}, {Shahaf}, {Soumagnac}, {Tal-Or}, {Santen}, {Zackay}, {Guttman}, {Rekhi}, {Townsend}, {Weinstein}, \& {Wold}}]{2024ApJ...964...74S}
{Shvartzvald}, Y., {Waxman}, E., {Gal-Yam}, A., {et~al.} 2024, \apj, 964, 74

\bibitem[{{Skilling}(1975)}]{Skilling.1975a}
{Skilling}, J. 1975, MNRAS, 172, 557

\bibitem[{{Smith}(2014)}]{Smi14}
{Smith}, N. 2014, \araa, 52, 487

\bibitem[{{Smith} {et~al.}(2009){Smith}, {Hinkle}, \& {Ryde}}]{SmiHinRyd09}
{Smith}, N., {Hinkle}, K.~H., \& {Ryde}, N. 2009, \aj, 137, 3558

\bibitem[{{Smith} {et~al.}(2011){Smith}, {Li}, {Filippenko}, \& {Chornock}}]{2011MNRAS.412.1522S}
{Smith}, N., {Li}, W., {Filippenko}, A.~V., \& {Chornock}, R. 2011, \mnras, 412, 1522

\bibitem[{{Soria} {et~al.}(2025){Soria}, {Russell}, {Wiston}, {Cheng}, {Margutti}, {Rose}, {Ryder}, \& {Terreran}}]{2025arXiv250201740S}
{Soria}, R., {Russell}, T.~D., {Wiston}, E., {et~al.} 2025, arXiv e-prints, arXiv:2502.01740

\bibitem[{{Sturner} {et~al.}(1997){Sturner}, {Skibo}, {Dermer}, \& {Mattox}}]{1997ApJ...490..619S}
{Sturner}, S.~J., {Skibo}, J.~G., {Dermer}, C.~D., \& {Mattox}, J.~R. 1997, \apj, 490, 619

\bibitem[{{Sushch} {et~al.}(2025){Sushch}, {Blasi}, \& {Brose}}]{2025arXiv250502523S}
{Sushch}, I., {Blasi}, P., \& {Brose}, R. 2025, arXiv e-prints, arXiv:2505.02523

\bibitem[{{Sushch} \& {van Soelen}(2017)}]{2017ApJ...837..175S}
{Sushch}, I. \& {van Soelen}, B. 2017, \apj, 837, 175

\bibitem[{{Taddia} {et~al.}(2020){Taddia}, {Stritzinger}, {Fransson}, {Brown}, {Contreras}, {Holmbo}, {Moriya}, {Phillips}, {Sollerman}, {Suntzeff}, {Ashall}, {Burns}, {Busta}, {Campillay}, {Castell{\'o}n}, {Corco}, {Di Mille}, {Gall}, {Gonz{\'a}lez}, {Hsiao}, {Morrell}, {Nyholm}, {Simon}, \& {Ser{\'o}n}}]{2020A&A...638A..92T}
{Taddia}, F., {Stritzinger}, M.~D., {Fransson}, C., {et~al.} 2020, \aap, 638, A92

\bibitem[{{Tatischeff}(2009)}]{2009A&A...499..191T}
{Tatischeff}, V. 2009, \aap, 499, 191

\bibitem[{{Telezhinsky} {et~al.}(2013){Telezhinsky}, {Dwarkadas}, \& {Pohl}}]{Telezhinsky.2013a}
{Telezhinsky}, I., {Dwarkadas}, V., \& {Pohl}, M. 2013, in {AAS/High Energy Astrophysics Division}, Vol.~13, {AAS/High Energy Astrophysics Division}, 127.16

\bibitem[{{Vernetto} \& {LHAASO Collaboration}(2016)}]{2016JPhCS.718e2043V}
{Vernetto}, S. \& {LHAASO Collaboration}. 2016, in Journal of Physics Conference Series, Vol. 718, Journal of Physics Conference Series, 052043

\bibitem[{{Wootten} \& {Thompson}(2009)}]{wootten09}
{Wootten}, A. \& {Thompson}, A.~R. 2009, IEEE Proceedings, 97, 1463

\bibitem[{{Xi} {et~al.}(2020){Xi}, {Liu}, {Wang}, {Yang}, {Yuan}, \& {Zhang}}]{2020ApJ...896L..33X}
{Xi}, S.-Q., {Liu}, R.-Y., {Wang}, X.-Y., {et~al.} 2020, \apjl, 896, L33

\bibitem[{{Yuan} {et~al.}(2018){Yuan}, {Liao}, {Xin}, {Li}, {Fan}, {Zhang}, {Hu}, \& {Bi}}]{2018ApJ...854L..18Y}
{Yuan}, Q., {Liao}, N.-H., {Xin}, Y.-L., {et~al.} 2018, \apjl, 854, L18

\bibitem[{{Zanardo} {et~al.}(2010){Zanardo}, {Staveley-Smith}, {Ball}, {Gaensler}, {Kesteven}, {Manchester}, {Ng}, {Tzioumis}, \& {Potter}}]{2010ApJ...710.1515Z}
{Zanardo}, G., {Staveley-Smith}, L., {Ball}, L., {et~al.} 2010, \apj, 710, 1515

\end{thebibliography}

\appendix
\section{Radiation and absorption processes}\label{sec:Ap_rad}

\subsection{Non-thermal emission}\label{sec:gamma-em}
The non-thermal emission of the accelerated particles is calculated in a spatially dependent manner, taking into account the local particle distribution and magnetic field strength in case of synchrotron emission or local plasma density in case of pion-decay emission. 
For the pion-decay emission, we follow the method described by \cite{2007APh....27..429H} whereas we use the cross-section data and gamma-ray production rates from \cite{2020APh...12302490B}.
The non-thermal synchrotron emission uses the standard approach presented, for instance, in \cite{1997ApJ...490..619S}, and accounts for the magnetic field component of the large-scale field as well as for the turbulent component self-amplified by the CRs.

\subsection{Thermal X-ray emission} \label{sec:Xray}
The thermal continuum X-ray emission is modeled using the formulations established by \cite{1999A&A...344..295H}. In alignment with the findings of \cite{2014MNRAS.440.1917D}, our analysis focuses exclusively on the emission originating from the supernova remnant's forward shock. This approach is justified by the early development of a dense, low-temperature shell around the reverse shock during the initial stages of supernova remnant evolution, particularly in the high-density environments considered in this study. Consequently, we omit any thermal emission from regions within $r<0.85\cdot R_\text{sh}$.

%We model the thermal continuum X-ray emission using the expressions derived by \cite{1999A&A...344..295H}. We follow the results of \cite{2014MNRAS.440.1917D} and consider only the emission from the forward shock of the remnant. It is suggested that a dense, low-temperature shell forms around the reverse shock at the early stages of the SNR evolution and high-density environments that are considered here. We thus exclude any thermal emission from regions inside of $r<0.85\cdot R_\text{sh}$.

\subsection{Secondary emission}\label{sec:Optical_method}
It can be expected that the shock-shell interaction also leads to a (re-)rise of the optical emission from the SNe, since a part of the thermal X-rays created by the shock-heated material gets absorbed in the dense CSM around the shocks and re-emitted also at optical wavelengths. A full radiation-transfer calculation is beyond the scope of this paper and hence we calculate the optical emission based on the luminosity of the absorbed X-rays ($L_\text{X-ray, absorbed}$).
\begin{align}
    L_\text{secondary} =  \epsilon (L_\text{X-ray,unabsorbed}-L_\text{X-ray, absorbed}),
\end{align}
where epsilon reflects the fact that only a fraction of the absorbed luminosity is re-processed. Here, we adopted $\epsilon=1$, meaning that our results have to be considered as upper limits.
%\is{IS: Should we just say that we treat it as an upper limit? and no estimate is needed then}
%\jm{[I would have thought it better to say $L_\text{optical} = \epsilon (L_\text{X-ray,unabsorbed}-L_\text{X-ray, absorbed})$, where $\epsilon<1$.  The absorbed luminosity is what emerges in X-rays after the rest of the emission has been re-processed to other wavelengths.  Only a fraction comes out in optical because a lot should appear in thermal IR from dust, and UV lines.]}
We then assume that the absorption is dominantly taking place close to the shock, so that the size of the emission region equals the current shock-radius. Hence, 
\begin{align}
    L_\text{secondary} &= \sigma 4 \pi R_\text{Sh}^2 T^4 , 
\end{align}
and calculate the emission temperature. We then use Planck's law to calculate the emission in the B-band (398nm-492nm), the R-band (591nm-727nm) and in UV\footnote{This corresponds to the wavelengths of the upcoming ULTRASAT mission \citealp{2024ApJ...964...74S}.} (230-290nm) and integrate to obtain the luminosity. %With the example parameters for the SN photospheres \citepalias[see][]{2022MNRAS.516..492B} we calculate the luminosity for the optical SN-emission itself.
%\jm{[I'm not sure we want to go to predicting the temperature, because much of the optical emission will be in line emission rather than continuum, unless the density is huge and we create a new photosphere for the SN.  I think we are on shaky ground if we try to predict the colour temperature.  Better to just scale the absorbed X-rays by some very uncertain factor $\epsilon$ and say this will be the level of the optical emission.]}
It has to be noted that our calculations for the secondary emission are not suitable for the LBV scenarios, as there the interaction of the shock with the dense material gives rise to strong radiative losses and re-heating of the plasma. Taking these into account is beyond the scope of this paper and we hence only obtained the luminosities for the RSG cases. 

\subsection{Free-free absorption}
Free-free absorption (FFA) is considered the primary mechanism attenuating radio emissions from SNRs situated in dense circumstellar environments \citep[and references therein]{2021ApJ...908...75B}. \cite{1967ApJ...147..471M} formulated an expression for the absorption coefficient, $\kappa$, associated with FFA
\begin{align}
    \kappa &= 3.3\cdot10^{-7}\left(\frac{n_e}{\text{cm}^{-3}}\right)^2\left(\frac{T_e}{10^4 \text{K}}\right)^{-1.35}\left(\frac{\nu}{\text{GHz}}\right)^{-2.1}\text{pc}^{-1},
\end{align}
Here, $n_e$ represents the electron density, $T_e$ the electron temperature, and $\nu$ the frequency of the emitted radiation. The optical depth, $\tau_\mathrm{ff}$, is determined by integrating $\kappa$ along the observer's line of sight
\begin{equation}
    \tau_\mathrm{ff} = \int_\mathrm{LoS}\kappa dx . 
\end{equation}
In our study, we compute the synchrotron emission arising from the non-thermal electron population and the total magnetic field within the remnant. The resulting flux is then adjusted to account for attenuation caused by the intervening material along the line of sight.

\subsection{X-ray absorption}\label{sec:xray_abs}
We incorporate the attenuation of X-ray emission outside the supernova remnants due to the dense surrounding shells. The optical depth is computed using the expression
\begin{align}
    \tau &= \int_x^{\text{R}_\text{Shell}}n_H(x^\prime)\sigma_E\text{d}x^\prime \text{ , }
\end{align}
where $n_H$ represents the hydrogen number density at a given position, and $\sigma_E$ is the energy-dependent absorption cross-section \citep{2019MNRAS.486.1094M,1983ApJ...270..119M}, given by
\begin{align}
    \sigma_E &= 2.27\cdot10^{-22}E_{keV}^{-2.485}\,\text{cm}^2 \text{ , }\label{eq:SigmaX}
\end{align}
with $E_{keV}$ denoting the photon energy in keV. This relation approximates the absorption losses caused by heavy element ions in a gas with roughly solar abundances. While this is a simplification—particularly as it neglects the effects of ionization triggered by the supernova flash—it serves as a practical approach for our model. Notably, previous studies suggest that up to  $2M_\odot$ of material can be ionized by the SN emission \citep{1996ApJ...464..924L}. In our calculations, we estimate the extent of this ionized region and consider only material beyond this shell to contribute to X-ray absorption. Despite this, our model may still overestimate the attenuation of thermal X-rays. However, our primary goal is to approximate the X-ray output and enable cross-wavelength comparisons, particularly concerning the timing of emission features.

%is the photon-energy in keV. Equation (\ref{eq:SigmaX}) approximates losses due to absorption by ions of heavy elements in a gas with approximately solar abundances. This is of course an over-simplification of the process since ionization plays an important role e.g. by the SN flash, which cannot be accounted for in our model. However, it was suggested in the literature, that up to $2M_\odot$ of material can be ionized by the SN emission \citep{1996ApJ...464..924L}. During the absorption calculation, we estimate the size of this ionized region and take only material outside the ionized shell into account as attenuating the X-ray emission. Still, we might over-estimate the absorption of thermal X-rays. However, the scope of this calculation is to get an idea of the thermal X-ray emission and allow comparisons to other wavelengths, especially with respect to the timing of the emission.

\subsection{$\gamma\gamma$-absorption}
\label{sec:gg_absorption}
The $\gamma$-ray photons produced by shock-accelerated CRs can experience attenuation through pair production when interacting with the intense photon fields of the SN photosphere. As demonstrated by \cite{2009A&A...499..191T}, this mechanism can substantially reduce the VHE $\gamma$-ray flux during the first year post-explosion. Subsequent studies have refined this model by incorporating the anisotropic nature of the photon field and accounting for light-travel time effects \citep{2020MNRAS.494.2760C}. Due to the computational demands of these calculations, \citet{2020MNRAS.494.2760C} assumed that emission originates solely from a thin shell. In \citetalias{2022MNRAS.516..492B}, we extended this approach by considering the spatial distribution of the emission region, though light-travel effects were neglected. These studies collectively indicate that VHE flux suppression is most pronounced within the first several tens to hundreds of days, varying with energy band, and is less severe than predictions based on isotropic interactions. Notably, for smooth stellar winds, this period coincides with the peak of unabsorbed $\gamma$-ray emission from SNRs.

Building upon the methodology of \citetalias{2022MNRAS.516..492B}, we incorporate the anisotropic characteristics of $\gamma\gamma$ interactions and the finite extent of the photosphere. We omit light-travel time effects, which are primarily significant for VHE attenuation within the initial 10 days. Our computation of optical depth encompasses the entire remnant, accounting for the spatial distribution of emitting particles by evaluating the optical depth at each location within a radius of $r_\mathrm{max} = 1.2R_\mathrm{sh}(t)$, where $R_\mathrm{sh}(t)$ denotes the shock radius at a given time.

To calculate $\gamma\gamma$ absorption, we adopt an approach analogous to that used for binary systems by \citet{2006A&A...451....9D}, further developed in \citet{2017ApJ...837..175S}. The optical depth $\tau_{\gamma\gamma}$ for a $\gamma$-ray photon with energy $E_\gamma = \epsilon_\gamma m_\mathrm{e} c^2$ traversing a distance $l$ is given by \citep{1967PhRv..155.1404G}
\begin{equation}
  \tau_{\gamma\gamma} = \int_0^l\,{\rm d}l \int_{4\pi}  {\rm d}\Omega \, \,(1-\mu) \int_{\frac{2}{\epsilon_\gamma (1-\mu)}}^\infty {\rm d} \epsilon \, n_\mathrm{ph}(\epsilon,\Omega) \sigma_{\gamma\gamma}(\epsilon,\epsilon_\gamma,\mu) . 
\label{eqn:tau_gamma}
\end{equation}
Here, $\mu = \cos \theta$, with $\theta$ representing the angle between the $\gamma$-ray photon and the low-energy target photon. The solid angle element is ${\rm d}\Omega = \sin{\theta^\prime}d\theta^\prime d\phi^\prime$, defined in a spherical coordinate system centered on the $\gamma$-ray photon, with the zenith aligned toward the photosphere's center. The target photon energy, normalized to the electron rest mass, is $\epsilon = h\nu/(m_\mathrm{e} c^2)$, and $n_{\rm ph}(\epsilon,\Omega)$ denotes the differential number density of target photons per unit solid angle.

The $\gamma\gamma$ pair production cross-section $\sigma_{\gamma\gamma}$ is expressed as \citep{1976tper.book.....J}
\begin{equation}
 \sigma_{\gamma\gamma}(\beta) = \frac{3}{16} \sigma_{\rm T} ( 1- \beta^2) \left[ (3-\beta^4) \ln \left(\frac{1+\beta}{1-\beta} \right) -2\beta (2-\beta^2)\right],
\end{equation}
with
\begin{equation}
 \beta = \sqrt{1-\frac{2}{\epsilon \epsilon_\gamma (1-\mu)}},
\end{equation}
where $\sigma_{\rm T}$ is the Thomson cross-section.

\begin{figure*}
\centering
\includegraphics[width=0.99\textwidth]{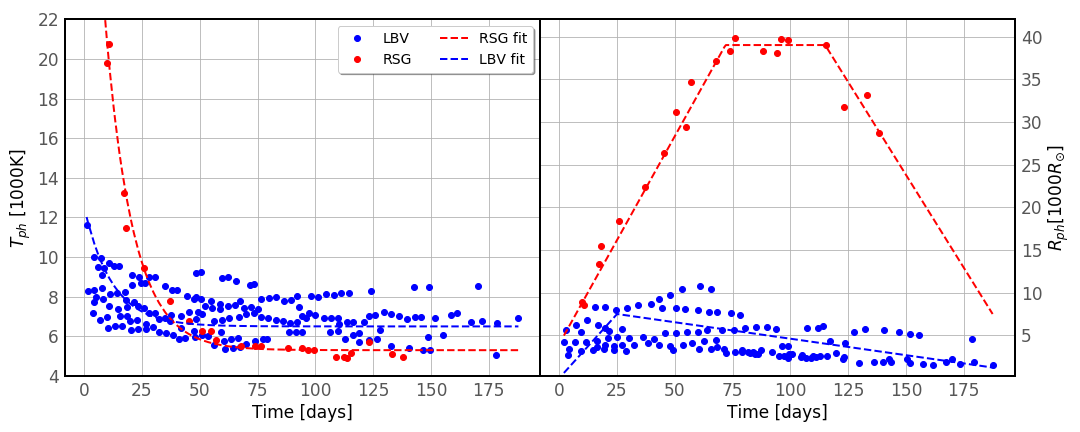}
\caption{Photosphere temperature (left) and radius (right) for Type-IIn/LBV (blue) and Type-IIP/RSG (red) supernovae. Data points are sourced from \protect\cite{2020A&A...638A..92T} and \protect\cite{10.1093/mnras/sty1634}. Figure adapted from \citetalias{2022MNRAS.516..492B}.}
\label{fig:Photospheres}
\end{figure*}

We derive the photospheric properties for Type-IIn and Type-IIP supernovae from observational data provided by \cite{2020A&A...638A..92T} and \cite{10.1093/mnras/sty1634}, respectively. The photospheric radius is modeled using piecewise linear fits, while the temperature evolution is described by an exponential decay. These fit functions are illustrated alongside the observational data in Figure \ref{fig:Photospheres}.

Assuming the photospheric photon field follows a blackbody distribution, the photon number density is given by
\begin{equation}
    n_\mathrm{ph}(\nu,\Omega) = \frac{2\nu^2}{c^3}\frac{1}{e^\frac{h\nu}{kT_\mathrm{ps}}-1}.
\end{equation}
Utilizing these photospheric parameters, we compute the opacity for photons emitted from various regions within the remnant, thereby obtaining the attenuated $\gamma$-ray spectra.

In \citetalias{2022MNRAS.516..492B}, absorption matrices were computed for select time points during the photosphere's evolution, with interpolation employed to estimate $\gamma$-ray attenuation at intermediate times. In this study, we enhance our numerical approach by implementing GPU acceleration, enabling the computation of absorption matrices at every output time step. This advancement eliminates the need for interpolation, thereby reducing numerical artifacts previously observed, such as those depicted in Figure 4 of \citetalias{2022MNRAS.516..492B}.

It is important to note that our assumptions on the SN-photospheres are crucial for the resulting lightcurves and can differ widely between different SN. Some Type-IIn SNe show photospheric emission for more then five years as in the case of SN 2017hcc \citep{2023A&A...669A..51M}. However, also in extreme cases, the \gaga-absorption will still fade considerably with time as $R_\text{ph}/R_\text{sh}$ will approach zero over time, resulting in small scattering angles and hence less and less absorption.

We plan to release the \gaga-absorption code freely in an upcoming publication.

\subsection{EBL-absorption}
In addition to the photon fields discussed in Section \ref{sec:gg_absorption}, the extragalactic background light (EBL) constitutes another significant contributor to the $\gamma\gamma$ absorption process for extragalactic sources \citep[see e.g.,][]{1966PhRvL..16..252G}.

To assess the detectability of distant Type-IIP and Type-IIn supernovae, we incorporate the effects of EBL-induced attenuation. We adopt the EBL model proposed by \cite{2010ApJ...712..238F} to characterize the spectral energy distribution of the EBL. Using this model, we compute the $\gamma\gamma$ optical depth of the Universe, denoted as $\tau_{\gamma\gamma}^{\mathrm{EBL}}$, for gamma-ray photons with energies corresponding to the lower and upper thresholds of various gamma-ray observatories, as a function of the source distance $d$. These calculations are performed utilizing the Astropy library \citep{astropy:2018}.

The average of the optical depths at the two energy thresholds is then used to determine the attenuation factor, which is applied to the gamma-ray flux as a multiplicative correction:
\begin{equation}
    F_\text{attenuated}=F_\text{intrinsic}\times \exp\left(-\tau_{\gamma\gamma}^\text{EBL}\right)
\end{equation}

This approach allows us to account for the reduction in observable gamma-ray flux due to interactions with the EBL when evaluating the prospects for detecting distant supernova events. In this work, the EBL-absorption becomes significantly more important as the shell-interactions can boost the gamma-ray emission significantly beyond the levels of the smooth winds we considered earlier.

\section{Detectability}\label{sec:Detectability}
We evaluated the detectability of the emission for various present and future gamma-ray observatories. 
\subsection{HE detectability}
\begin{figure}[h!]
    \centering
  \includegraphics[width=0.45\textwidth]{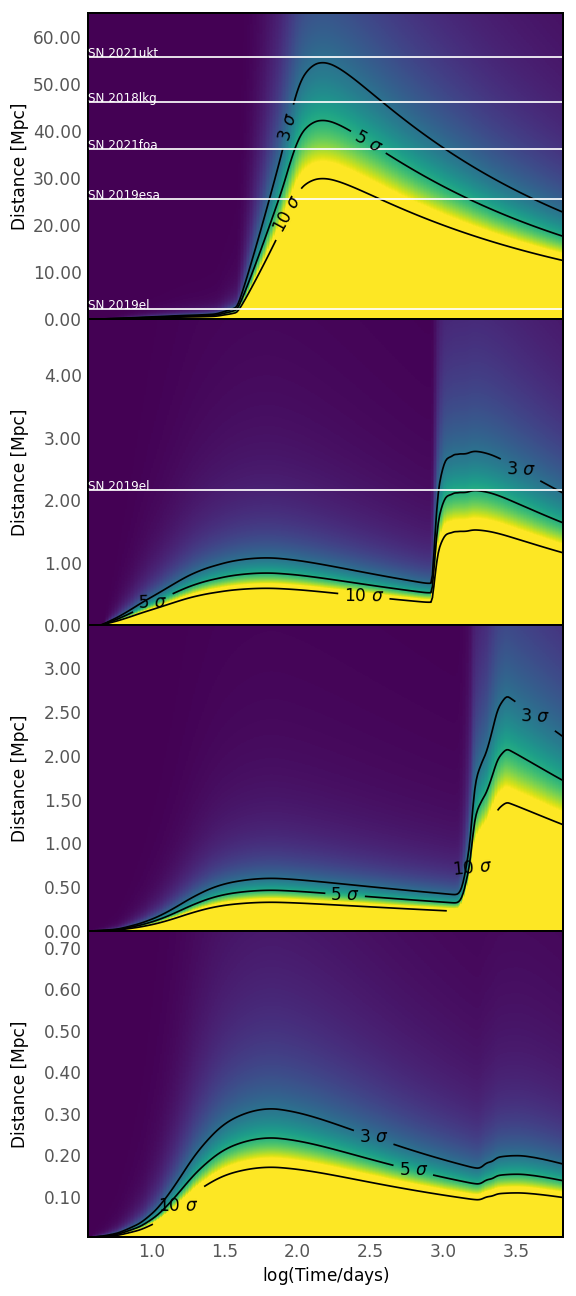}
    \caption{eASTROGAM detectability for \rb{LBV 0.1, LBV 2.0, RSG dense shell and RSG dilute shell} scenarios from top to bottom. White lines indicate detected SNe explosions at their inferred distance.}
%    \caption{Detectability of nearby SNe by the survey-like experiments \emph{Fermi-LAT} (1-300GeV), \emph{HAWC} (2-20TeV) and \emph{LHAASO} (40-300TeV). Shown is the culminated significance for a SNR observed since explosion. The linear colorscale spans from $0$ to $10\sigma$.}
    \label{fig:Astrogam}
\end{figure}

The HE gamma-ray domain is less affected by \gaga absorption that observatories operation at higher energies. At the moment, the only relevant instrument in the HE gamma-ray domain is Fermi-LAT, where we already discussed the detectability in section \ref{sec:gamma_vis} and figure \ref{fig:Fermi}. The detectability for Fermi is evaluated between $1-300\,$GeV. 

eASTROGAMM is proposed to launch in the next decade and to observe the gamma-ray sky at slightly lower energies - here we used $100-1000\,$MeV - than Fermi-LAT with a compatible sensitivity but a considerably better angular resolution \citep{2017ExA....44...25D}. Figure \ref{fig:Astrogam} shows the detection horizon for eASTROGAM, which is compatible with Fermi-LAT's.

\subsection{VHE-detectability}
In the VHE-domain, the currently relevant instruments are the IACTs H.E.S.S., VERITAS and Magic, which all have a comparable sensitivity at higher energies and the water Cherenkov telescope HAWC.

\begin{figure}[h!]
    \centering
    \includegraphics[width=0.45\textwidth]{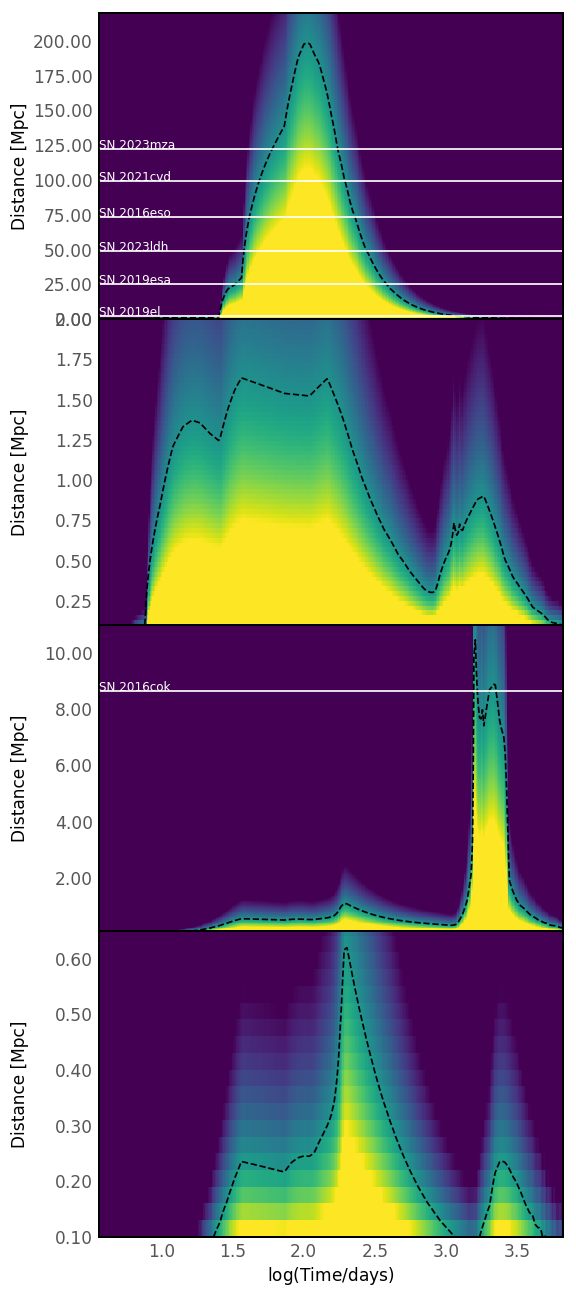}
    \caption{Same as figure \ref{fig:Astrogam} but for the H.E.S.S. instrument.}
    \label{fig:hess}
\end{figure}

\begin{figure}[h!]
    \centering
    \includegraphics[width=0.45\textwidth]{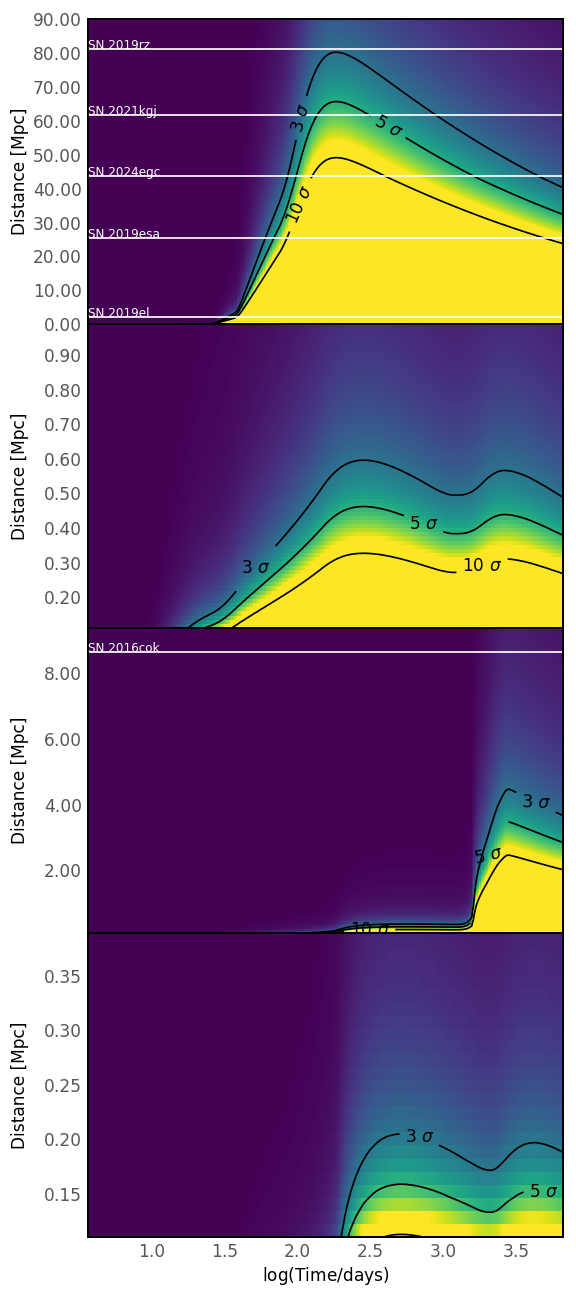}
    \caption{Same as figure \ref{fig:Astrogam} but for the HAWC instrument.}
    \label{fig:hawc}
\end{figure}

Here, we only used the H.E.S.S.-sensitivity as a benchmark for the IACTs (see figure \ref{fig:hess}) and integrated the spectrum between $1-10\,$TeV. For HAWC, which is more sensitive at higher energies, we used $2-20\,$TeV (see figure \ref{fig:hawc}).

\begin{figure}[h!]
    \centering    
    \includegraphics[width=0.45\textwidth]{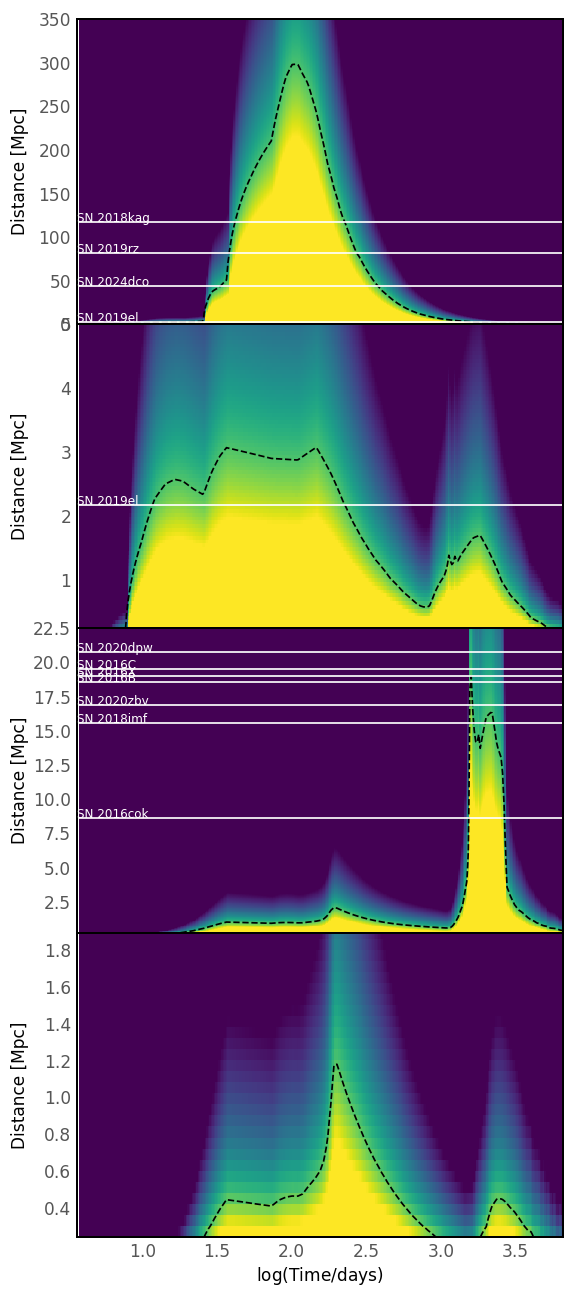}
    \caption{Same as figure \ref{fig:Astrogam} but for the the planned CTAO-South observatory.}
%    \caption{Detectability of nearby SNe by the IACTs \emph{H.E.S.S.}\ and \emph{CTAO-South}. The logarithmic color scale shows flux as a function of time post-explosion and of distance to the SN, where lighter colors denote a higher flux. The dashed lines indicate the flux needed for a $5\sigma$ detection within 50hrs for both instruments. The color-scale spans from 1/5th to five-times the detection-flux in all cases. Note, middle peaks that can be seen in all four panels is the numeric artifact (see the main text and Appendix for details).}
   \label{fig:cta}
\end{figure}

In the future, CTAO will come into operation in the VHE domain and provide a superior sensitivity compared to current-generation instruments. For CTAO, we integrated or emission spectra between $1-20\,$TeV (see figure \ref{fig:cta}).

%HESS(1.5Mpc; 75kpc), VERITAS, Magic, HAWC(1.5Mpc, 75kpc), CTAS(5Mpc, 0.2Mpc)

\subsection{UHE-detectability}
\begin{figure}[h!]
    \centering
    \includegraphics[width=0.45\textwidth]{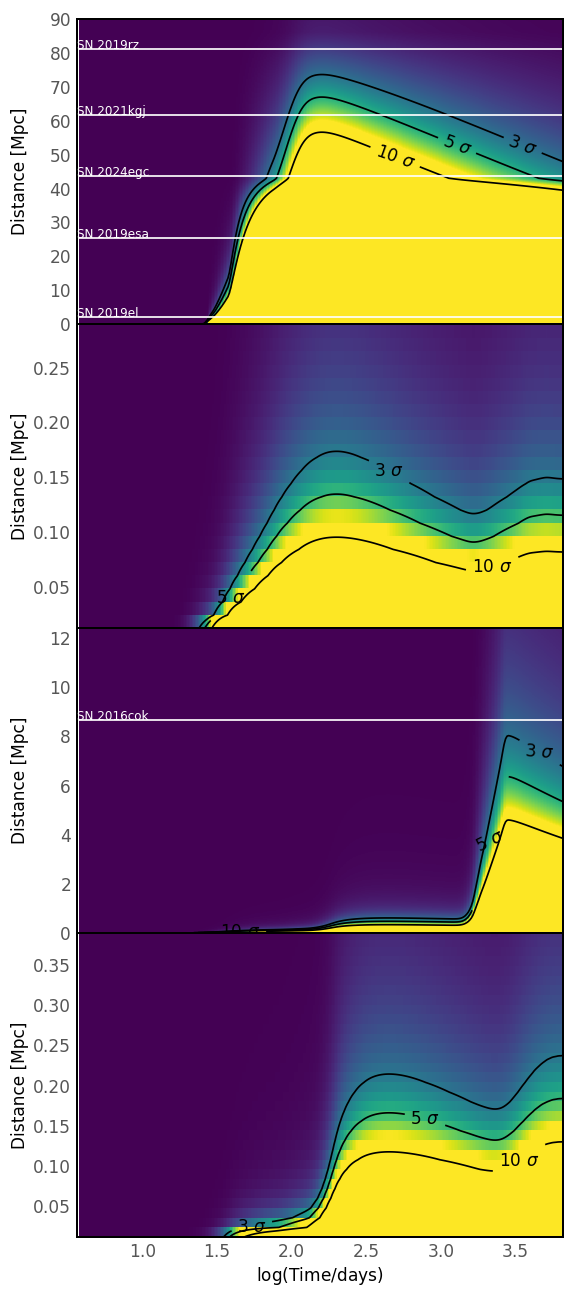}
    \caption{Same as figure \ref{fig:Astrogam} but for the LHAASO observatory.}
%    \caption{Detectability of nearby SNe by the IACTs \emph{H.E.S.S.}\ and \emph{CTAO-South}. The logarithmic color scale shows flux as a function of time post-explosion and of distance to the SN, where lighter colors denote a higher flux. The dashed lines indicate the flux needed for a $5\sigma$ detection within 50hrs for both instruments. The color-scale spans from 1/5th to five-times the detection-flux in all cases. Note, middle peaks that can be seen in all four panels is the numeric artifact (see the main text and Appendix for details).}
    \label{fig:lhaaso}
\end{figure}

LHAASO provides us with a new window to the gamma-ray sky at energies of up to a few PeV. For the LHAASO-sensitivity plot, we integrated in the energy range of $40-300\,$TeV (see figure \ref{fig:lhaaso}).

%LHAASO(3Mpc, 0.2Mpc) SWIGO(?)
%\emph{LHAASO} recently opened the window to ultra-high-energy (UHE) $\gamma$-ray astronomy \citep{2021ChPhC..45b5002A}.
%We investigated the capability of \emph{LHAASO} to discover $\gamma$-rays of nearby SNRs similarly to the method for \emph{Fermi-LAT} and \emph{HAWC} (\ref{fig:Survey}, panel e) and f)) \citep{2016JPhCS.718e2043V}. It turns out that the detection threshold is $1\,$Mpc and $0.04\,$Mpc for a Type-IIn and Type-IIP SNRs respectively, giving it similar capabilities to \emph{CTAO South}. However, the differential sensitivity of \emph{LHAASO} significantly increases beyond $100\,$TeV of $\gamma$-ray energy. In our models, the $\gamma$-ray spectra cut-off around $75\,$TeV, thus not benefiting from this high-sensitivity region. It can be expected, however, that \emph{LHAASO} is very sensitive to any SNR that accelerates CRs to higher energies, hence exceptional objects that expand in highly magnetized ambient media or feature a higher explosion energy, like superluminous SNe.  

\end{document}